\documentclass{article}

\usepackage{PRIMEarxiv}

\usepackage[utf8]{inputenc} 
\usepackage[T1]{fontenc}    
\usepackage{hyperref}       
\usepackage{url}            
\usepackage{booktabs}       
\usepackage{amsfonts}       
\usepackage{nicefrac}       
\usepackage{microtype}      
\usepackage{lipsum}
\usepackage{fancyhdr}       
\usepackage{graphicx}       
\graphicspath{{figures/}}     
\usepackage{algorithm} 
\usepackage{algpseudocode} 
\usepackage{amsmath}
\usepackage{soul,color}
\usepackage{xspace,amstext}
\usepackage[table,xcdraw]{xcolor}
\usepackage{float}
\usepackage{graphicx}
\usepackage{caption}
\usepackage{subcaption}
\usepackage{titlesec}
\usepackage{multirow}
\usepackage{amsmath}
\usepackage{soul,color}
\usepackage[section]{placeins}

\title{TCSP: a Template based crystal structure prediction algorithm and web server for materials discovery
}

\author{
  Lai Wei \\
 Department of Computer Science and Engineering\\
  University of South Carolina\\
  Columbia, SC 29201 \\
  \And
 Nihang Fu, Edirisuriya M. D. Siriwardane \\
 Department of Computer Science and Engineering\\
  University of South Carolina\\
  Columbia, SC 29201 \\
  \And
  Wenhui Yang\\
 School of Mechanical Engineering\\
 Guizhou University\\
  Guiyang, China 550050 \\  
    \And
  Sadman Sadeed Omee, Rongzhi Dong, Rui Xin\\
 Department of Computer Science and Engineering\\
  University of South Carolina\\
  Columbia, SC 29201 \\
   \And
 Jianjun Hu *\\
 Department of Computer Science and Engineering\\
  University of South Carolina\\
  Columbia, SC 29201 \\
  \texttt{jianjunh@cse.sc.edu} \\
}

\begin{document}
\maketitle

\begin{abstract}

Fast and accurate crystal structure prediction (CSP) algorithms and web servers are highly desirable for exploring and discovering new materials out of the infinite design space. However, currently, the computationally expensive first principle calculation based crystal structure prediction algorithms are applicable to relatively small systems and are out of reach of most materials researchers due to the requirement of high computing resources or the software cost related to ab initio code such as VASP. Several computational teams have used an element substitution approach for generating or predicting new structures, but usually in an ad hoc way. Here we develop a template based crystal structure prediction algorithm (TCSP) and its companion web server, which makes this tool to be accessible to all materials researchers. Our algorithm uses elemental/chemical similarity and oxidation states to guide the selection of template structures and then rank them based on the substitution compatibility and can return multiple predictions with ranking scores in a few minutes. Benchmark study on the ~98,290 formulas of the Materials Project database using leave-one-out evaluation shows that our algorithm can achieve high accuracy (for 13,145 target structures, TCSP predicted their structures with RMSD < 0.1) for a large portion of the formulas. We have also used TCSP to discover new materials of the Ga-B-N system showing its potential for high-throughput materials discovery. Our user-friendly web app TCSP can be accessed freely at \url{www.materialsatlas.org/crystalstructure} on our MaterialsAtlas.org web app platform.

\end{abstract}


\section{Introduction}


Crystal structure prediction is increasingly becoming one of the most effective approaches for discovering new functional materials \cite{oganov2019structure} due to the ease to obtain new compositions either by enumeration \cite{davies2016computational}, heuristic knowledge, or the latest deep learning-based generative machine learning models \cite{dan2020generative}. While the peer protein structure prediction problem has been recently almost solved by the deep learning-based AlphaFold and RossettaFold algorithms, the crystal structure prediction problem remains elusive for a majority of categories of compositions. There are mainly three types of crystal structure prediction approaches including the ab initio based global optimization \cite{glass2006uspex, wang2015materials,liang2020cryspnet, demir2021ffcasp,falls2020xtalopt}, machine learning-based prediction \cite{hu2021alphacrystal}, and template-based elemental substitution \cite{hautier2011data}. The first approach instead depends on computationally expensive DFT calculations and is applicable to only small chemical systems. The second approach is inspired by the AlphaFold family of deep learning algorithms \cite{senior2020improved,jumper2021highly}, but is only at the early stage of development. The last template-based CSP methods are the most widely used and easiest to implement. Even though this method cannot predict crystal structures of new prototypes, recent deep generative models can discover new prototype materials which can partially address this issue \cite{zhao2021high}. In a pioneering work \cite{hautier2011data}, Hautier et al. proposed a data mining-based approach to identify the probabilities for different pairs of ionic substitutions, which can be applied to any prototype structures to generate new structures or used to select templates for template-based crystal structure prediction. However, despite the wide usage of template-based CSP methods, there are many different ways to implement, and there is no working web app/server that is user-friendly to make it accessible to all materials scientists (The structure predictor of the Materials Project \cite{ceder2010materials} website was available before but is not functional now). 

Here we propose a fast and user-friendly template-based crystal structure prediction algorithm and related companion web server for broad adoption of crystal structure prediction in the daily life of materials science. Our algorithm TCSP is based on the careful selection of template structures based on chemical formula similarity and the match of oxidation states using an exhaustive enumeration strategy. Our predicted structures can be optimized by DFT or machine learning-based structure relaxation. By using seven case studies, we have shown that our user-friendly and fast crystal structure prediction web server has high prediction performance when appropriate templates are available. We also apply our TCSP algorithm to predict the structures of all 98,290 formulas using leave-one-out evaluation and have achieved good performance for a large portion of the targets: more than 13,145 target structures have been found with maximum RMSD less than 0.1. The good performance of this high-throughput crystal structure prediction shows that the template/prototype-based element substitution CSP approach has big potential in exploratory materials discovery. With the development of large scale prototype databases \cite{su2017construction,hicks2021aflow} and their applications in the generative design of new crystals \cite{bushlanov2019topology,sorkun2020artificial}, the performance of our template-based crystal structure prediction algorithm can be further improved.

\section{Method}
\label{sec:headings}

\subsection{Template based crystal structure prediction algorithm}

\begin{figure}[ht]
  \centering
  \includegraphics[width=0.45\linewidth]{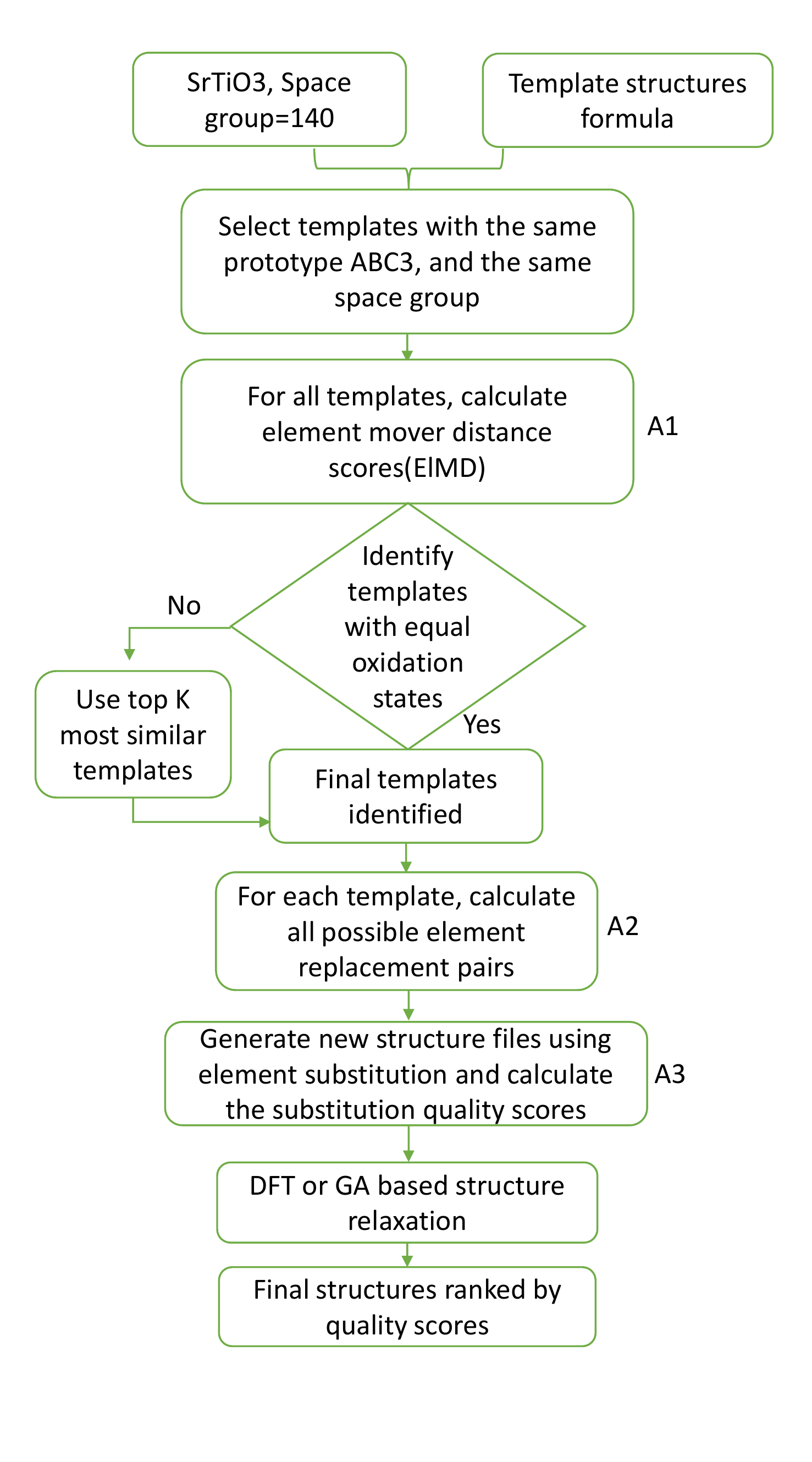}
  \caption{Flowchart of our TCSP, a template based crystal structure prediction algorithm. The space group specification is optional. }
  \label{fig:predict_algorithm}
\end{figure}

Our template-based crystal structure prediction algorithm is illustrated in Figure~\ref{fig:predict_algorithm}. Given an input formula (e.g., SrTiO\textsubscript{3}) with which can be specified by the user or predicted by algorithms \cite{zhao2020machine,li2021composition} or without space group. We then search structure templates with the same prototype (sometimes called an anonymous formula)  (e.g., ABC\textsubscript{3}) and the same space group if specified. This step may retrieve too many matched templates, so we use Module A1, an Element's mover distance, to measure the composition similarity between the query formula and the compositions of all the template structures, which are then ranked by ascending order. We then pick the top K structures as template candidates with the smallest composition distances. For each of the candidate templates, we use the Pymatgen package to estimate its oxidation states and compare them to those of the query formula. If we find templates with identical oxidation states, we then add them to the final template list. If no such templates are found, we then neglect the oxidation match requirements and directly add them as final templates. The next step is to determine all the possible element substitution pairs between the query formula and the template formula using the algorithm described in Algorithm 2. Next, we will pick the template structure files and replace the elements according to the pair arrangements found by Algorithm 2. A replacement quality score is also calculated for each such element substitution arrangement using the procedure as described in Module 3. The resulting structures will then be subject to DFT or machine learning-based structure relaxation, which can be further used to calculate formation energy, e-above-hull energy, and phonon dispersion for validation. 


\paragraph{Module A1: Element mover's distance for formula similarity calculation}
We use the Element's Mover Distance measure $ElMD$ \cite{hargreaves2020earth} to select most similar template structures. ElMD is a metric that allows measuring the chemical similarity of two formulas in an explainable fashion. The EMD is computed between two compositions from the ratio of each of the elements and the absolute distance between the elements on the modified Pettifor scale (several other element similarities can also be used such as Mendeleev, Petti, Atomic, Mod\_petti, Oliynyk, Oliynyk\_sc, Jarvis, Jvarvis\_sc, magpie,magpie\_sc, CGCNN, Elemnet, mat2vec, Matscholar, megnet16, random). This metric shows clear strength at distinguishing compounds. It is shown that ElMD distances have greater alignment with chemical understanding than the Euclidean distance. The ElMD is defined in formula (1) below:


\begin{equation}
ElM D(X, Y)=\min \sum_{i=1}^{m} \sum_{j=1}^{n} q_{i j}\left|p_{i}-p_{j}\right|
\textrm{, subject to} \quad q_{i j} \geq 0 \quad \textrm{for ${\forall}$ } i, j
\end{equation}

\begin{equation}
\sum_{j=1}^{n} q_{i j} \leq x_{i}  \textrm{, for ${\forall}$ } 1 \leq i \leq m
\end{equation}

\begin{equation}
\sum_{i=1}^{m} q_{i j} \leq y_{j} \textrm{,  for ${\forall}$ } 1 \leq j \leq n
\end{equation}

\begin{equation}
\sum_{i=1}^{m} \sum_{j=1}^{n} q_{i j}=1
\end{equation}

where the distance is first calculated by matching and pairing each of the $m$ elements in a vector, $X$, to its most similar unmatched partner in the $n$ elements of a second vector $Y$, until all have been paired. The quantity matched, $q$, from the $i$-th element of $X$, to the $j$-th element of $Y$, is given by $q_{ij}$.

\paragraph{Algorithm A2: element replacement pair enumeration algorithm:} 

This algorithm is used to enumerate all possible element replacement strategies between two pairs of formulas. 

\begin{algorithm}[h]
    \caption*{Algorithm A2: Element replacement pair enumeration algorithm}
    \begin{algorithmic}[1]
        \State Given two formula X,Y, calculate theirs oxidation states then get the statelist and elementlist.
        
        \If{$stateList_x == stateList_y$}
        
        {Replace element pairs}
        \Else{\State Create elementGroupList to represent and distinguish elements of equal state
            \For {$i = 0,1,\ldots$}
                \If{$i == 0$}
                    \State $elementGroup = elementList_y[i]$;
                \ElsIf {($stateList_y[i] == stateList_y[i-1]$)}
                \State append $elementList_y[i]$ to $elementGroup$;
                \Else
                    \State append $elementList_y[i]$ to $elementGroup$;
                \EndIf 
            \EndFor
            \State Create $pre\_patterns$ to represent permutation and combination of elements in $elementGroupList$
            
            \For{$element =0,1,\ldots,j$}
                \If{$j = 0$}
                    \State $pre\_patterns$ = permutations of $elementGroupList[j]$;
                \Else
                    \State $patterns$ =permutations of $elementGroupList[j]$;
                    \For{$p1 = 0,1,\ldots,m$} 
                        \For{$p2 = 0,1,\ldots,n$}
                            \State append $p1,p2$ to $new\_patterns$;
                        \EndFor
                        \State $pre\_patterns = new_patterns$;
                    \EndFor
                \EndIf 
            \EndFor
          
            \State If the element in $pre\_patterns$ is different from the element in $elementList_x$, then replace the element

        \EndIf
}
    \end{algorithmic} 
\end{algorithm}

\paragraph{Module A3: element substitution scoring function:}
To differentiate the resulting structures from different element substitution arrangements between the query formula and the template structure and rank the final output structures from different templates, we use the ElMD distance in Module A1 to calculate the similarity score between each pair of substitution elements for a given query formula and the final structure. Then we sum up these similarity scores for all the element substitution pairs and use them to calculate the quality scores of final structures. 

\begin{equation}
S_{es}= ElMD(e1,e2), S_{r}= \sum_{i=1}^{p}  S^{i}_{es}
\end{equation}

Where $S_{es}$ measures the element matching quality between two substitution elements $e1$ and $e2$. $S_{r}$ is the replacement distance score for a given element substitution arrangement of a given query formula and template structure, which is equal to the sum of all the element similarity scores of all substitution element pairs in the element substitution arrangement.

The final quality of the generated structures is then measured by the $S_r$ scores, with lower scores corresponding to higher quality. 



\paragraph{DFT or Genetic algorithm based structure relaxation:}

As with all predicted crystal structures, they usually need a fine-tuning or relaxation step to adjust the local atomic coordinates either using the DFT-based structural relaxation method or the recently developed machine learning and optimization-based relaxation approach \cite{zuo2021accelerating} which is much faster than the DFT approach. In this study, we used the DFT approach for evaluation purposes.

\FloatBarrier

\subsection{User interface of our web server}

\begin{figure}[ht]
  \centering
  \includegraphics[width=0.75\linewidth]{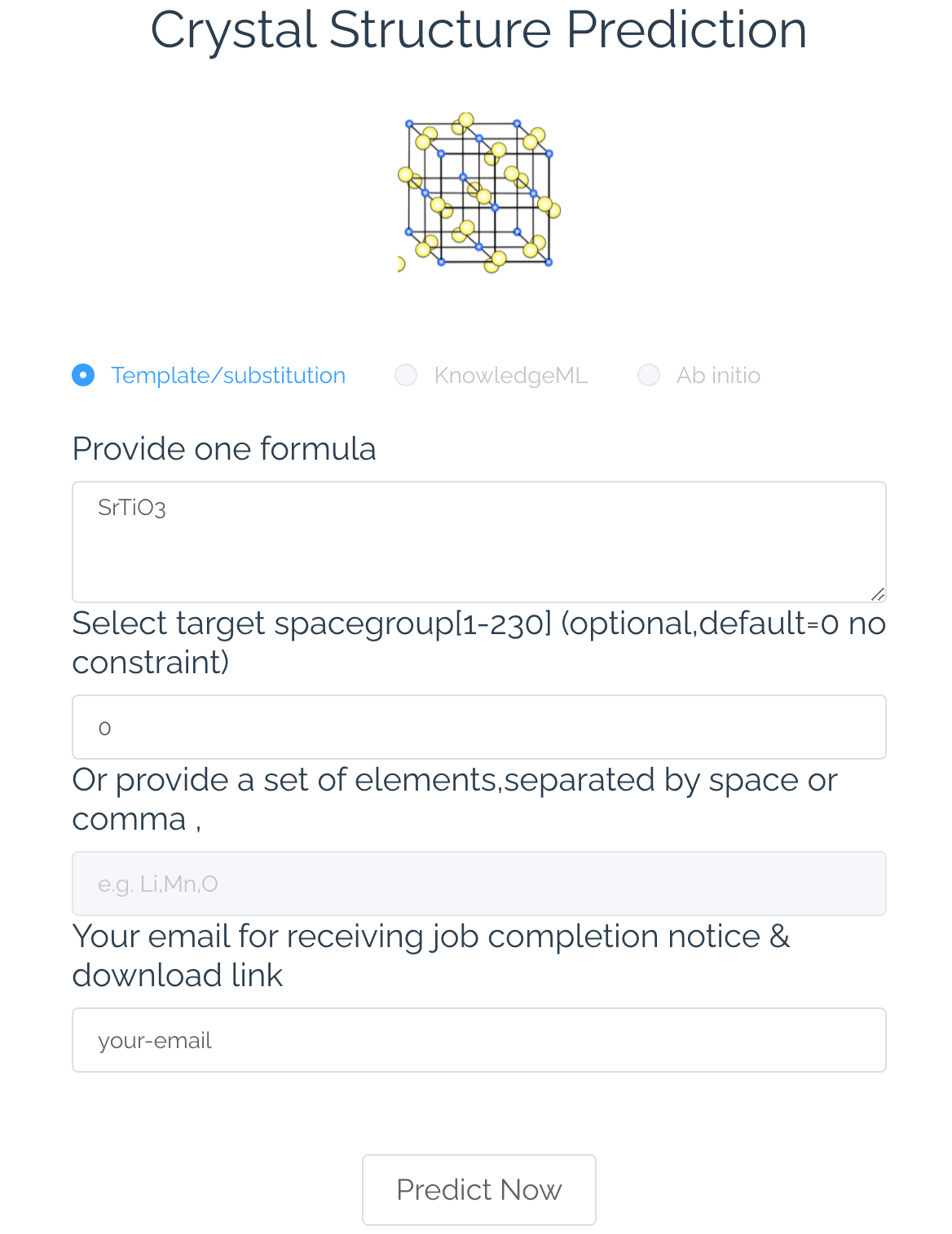}
  \caption{User interface of our TCSP web app for crystal structure prediction.}
  \label{fig:ui}
\end{figure}

Our template-based crystal structure prediction server has a user-friendly web interface, as shown in Figure\ref{fig:ui}. Each time, a user can just put in a formula/composition, and then the target space group number from 1 to 230 can be set or just assigned to 0 to allow a template with any space group. Then the user types in their email for receiving the job completion notification email with a downloadable URL link for the predicted structures. After a few minutes, an email will be sent to the user with the download URL link for the predicted structures. After the zipped result file is downloaded and unzipped, the user can go into the folder and click the Name column to sort the files by filename. Then it shows several key files: (1) $results.txt$, which shows the template similarity scores, the templates with compatible oxidation states, and the element replace pairs for each template. (2) $similar\_formulas.csv$ file shows the distance scores of all templates to the query formula; (3) $TemplateCandiates.csv$ shows the Materials Project IDs of the selected templates. (4) all the remaining $cif$ files are predicted, which are sorted by their replacement quality score (the number before \_mp of the filename), which is better when the number is smaller. However, it is strongly suggested to validate a couple of top-scored candidate structures as the candidate structure with the top quality score is not always the best one.

\subsection{DFT validation of predicted structures}

The first principle calculations based on the density functional theory (DFT) are carried out using the  Vienna \textit{ab initio} simulation package (VASP) \cite{Vasp1,Vasp2,Vasp3,Vasp4}.  The projected augmented wave (PAW) pseudopotentials, where 520 eV plane-wave cutoff energy, were used to treat the electron-ion interactions \cite{PAW1, PAW2}. The exchange-correlation functional was considered with the generalized gradient approximation (GGA) based on the Perdew-Burke-Ernzerhof (PBE) method \cite{GGA1, GGA2}. The energy convergence criterion was set as 10$^{-5}$ eV, while the atomic positions were optimized with the force convergence criterion of 10$^{-2}$ eV/{\AA}. The Brillouin zone integration for the unit cells was computed using the $\Gamma$-centered  Monkhorst-Pack $k$-meshes. The Formation energies (in eV/atom) of several materials were determined based on the expression in  Eq.~\ref{eq:form}, where $E[\mathrm{Material}]$ is the total energy per unit formula of the considered structure, $E[\textrm{A}_i]$ is the energy of $i^\mathrm{th}$ element of the material, $x_i$ indicates the number of A$_i$ atoms in a unit formula, and $n$ is the total number of atoms in a unit formula($n=\sum_i x_i$).

\begin{equation}
    E_{\mathrm{form}} =\frac{1}{n}(E[\mathrm{Material}] - \sum_i x_i E[\textrm{A}_i])
    \label{eq:form}
\end{equation}

\subsection{Evaluation criteria}


To evaluate the reconstruction performance of  algorithm, we define the root mean square distance (RMSD) and mean absolute error (MAE) of two structures as below:
\begin{equation}
    \begin{aligned}
\mathrm{RMSD}(\mathbf{v}, \mathbf{w}) &=\sqrt{\frac{1}{n} \sum_{i=1}^{n}\left\|v_{i}-w_{i}\right\|^{2}} \\
&=\sqrt{\frac{1}{n} \sum_{i=1}^{n}\left(\left(v_{i x}-w_{i x}\right)^{2}+\left(v_{i y}-w_{i y}\right)^{2}+\left(v_{i z}-w_{i z}\right)^{2}\right)}
\end{aligned}
\end{equation}

\begin{equation}
        \begin{aligned}
\mathrm{MAE}(\mathbf{v}, \mathbf{w}) &=\frac{1}{n} \sum_{i=1}^{n}\left\|v_{i}-w_{i}\right\| \\
&=\frac{1}{n} \sum_{i=1}^{n}\left(\|v_{i x}-w_{i x}\|+\|v_{i y}-w_{i y}\|+\|v_{i z}-w_{i z}\|\right)
\end{aligned}
\end{equation}

where $n$ is the number of independent atoms in the target crystal structure. For symmetrized CIF structures, $n$ is the number of independent atoms of the set of Wyckoff equivalent positions. For regular CIF structures, it is the total number of atoms in the compared structure. $v_i$ and $w_i$ are the corresponding atoms in the predicted crystal and the target crystal structure.

\section{Results}
\label{sec:others}

\subsection{Dataset}

We used more than 130,000 crystal structures deposited in the Materials Project database as our template sources. We picked seven test materials including SrTiO\textsubscript{3} (mp-5229), Ni\textsubscript{3}S\textsubscript{4} (mp-1050), NiS\textsubscript{2} (mp-849059), GaBN\textsubscript{2} (mp-1007823), GaB\textsubscript{3}N\textsubscript{4} (mp-1019740), GaB\textsubscript{2}N\textsubscript{3} (mp-1245554), and Ga\textsubscript{3}BN\textsubscript{4} (mp-1019743). We then ran our algorithm and checked if it can predict the correct structures that match the target structures.

\subsection{ Performance of template based crystal structure prediction}

\begin{figure}[ht!] 
 \begin{subfigure}[t]{0.33\textwidth}
        \includegraphics[width=\textwidth]{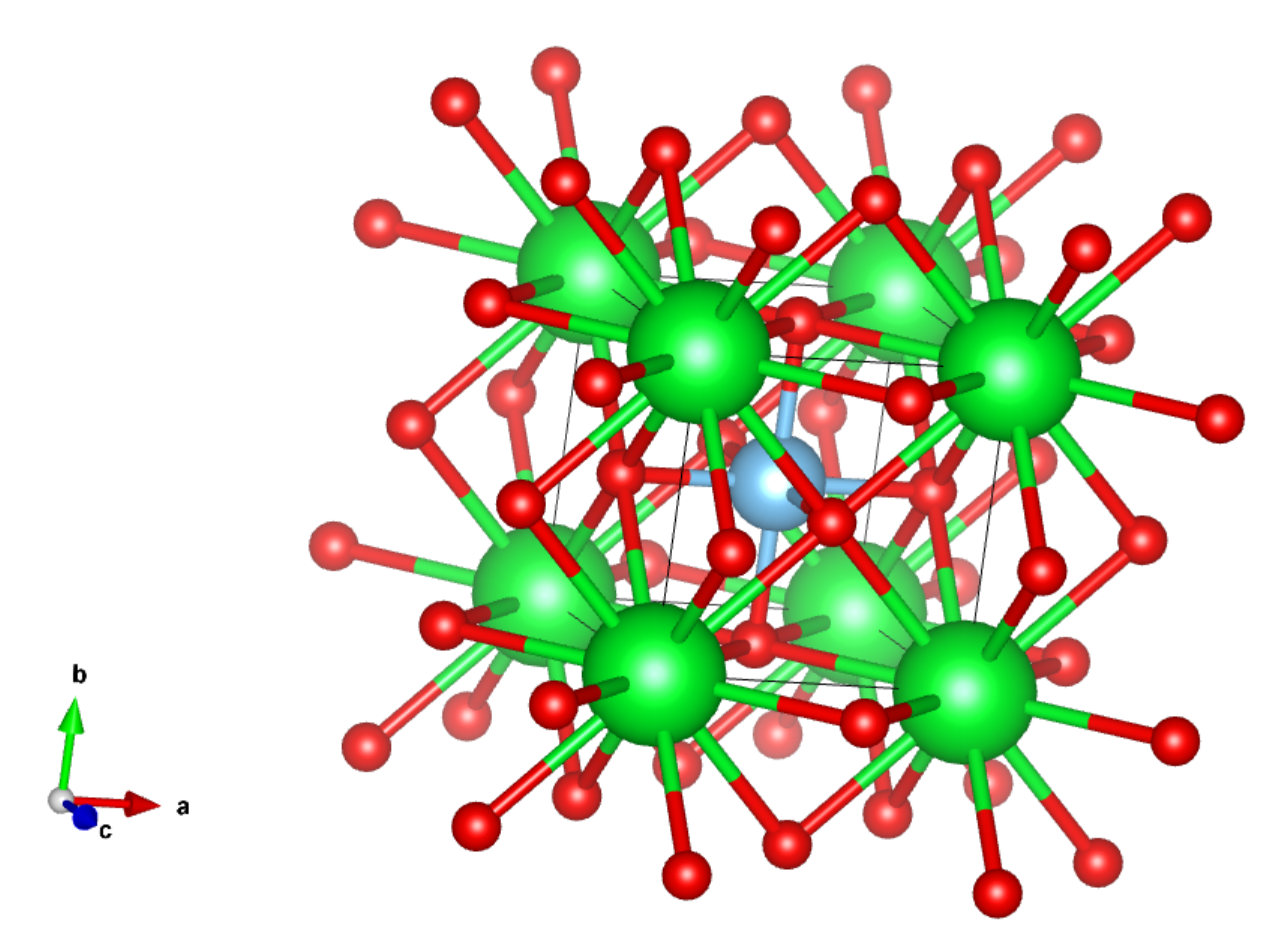}
        \caption{SrTiO\textsubscript{3}(Target)}
        \vspace{-3pt}
        \label{fig:SrTiO3_target}
    \end{subfigure}
    \begin{subfigure}[t]{0.33\textwidth}
        \includegraphics[width=\textwidth]{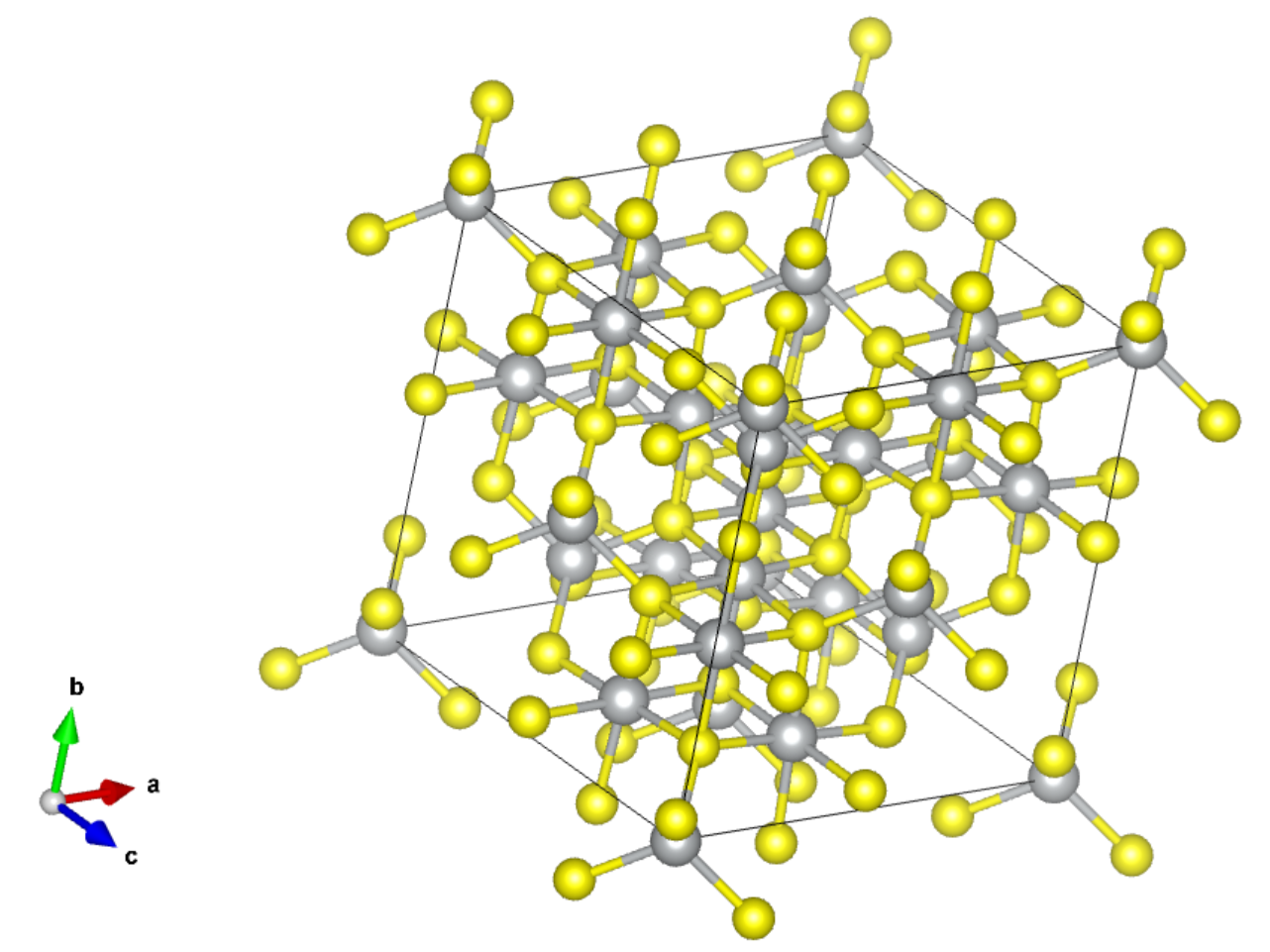}
        \caption{Ni\textsubscript{3}S\textsubscript{4}(Target)}
        \vspace{-3pt}
        \label{fig:Ni3S4_target}
    \end{subfigure}    
    \begin{subfigure}[t]{0.33\textwidth}
        \includegraphics[width=\textwidth]{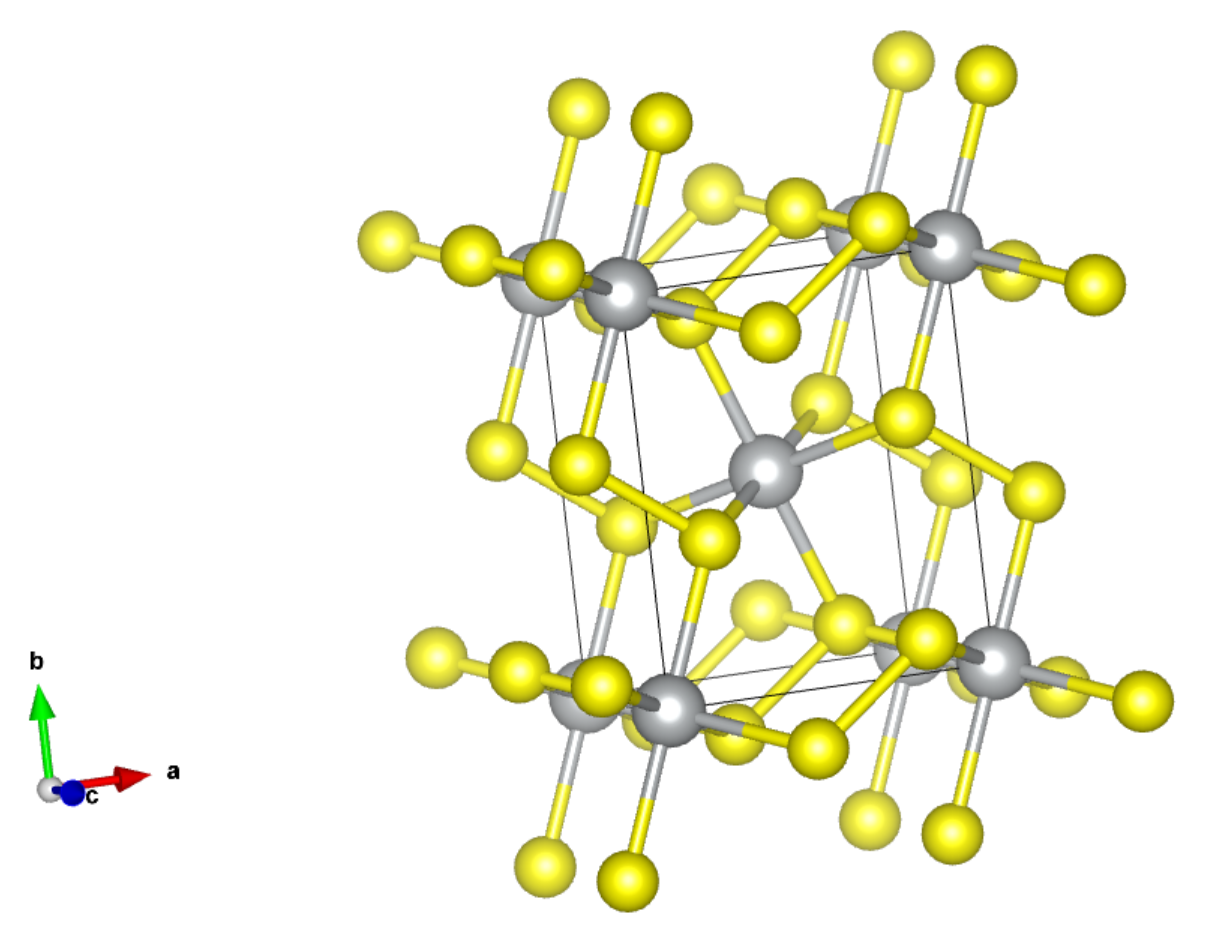}
        \caption{NiS\textsubscript{2}(Target)}
        \vspace{-3pt}
        \label{fig:NiS2_target}
    \end{subfigure}    
    \begin{subfigure}[t]{0.33\textwidth}
        \includegraphics[width=\textwidth]{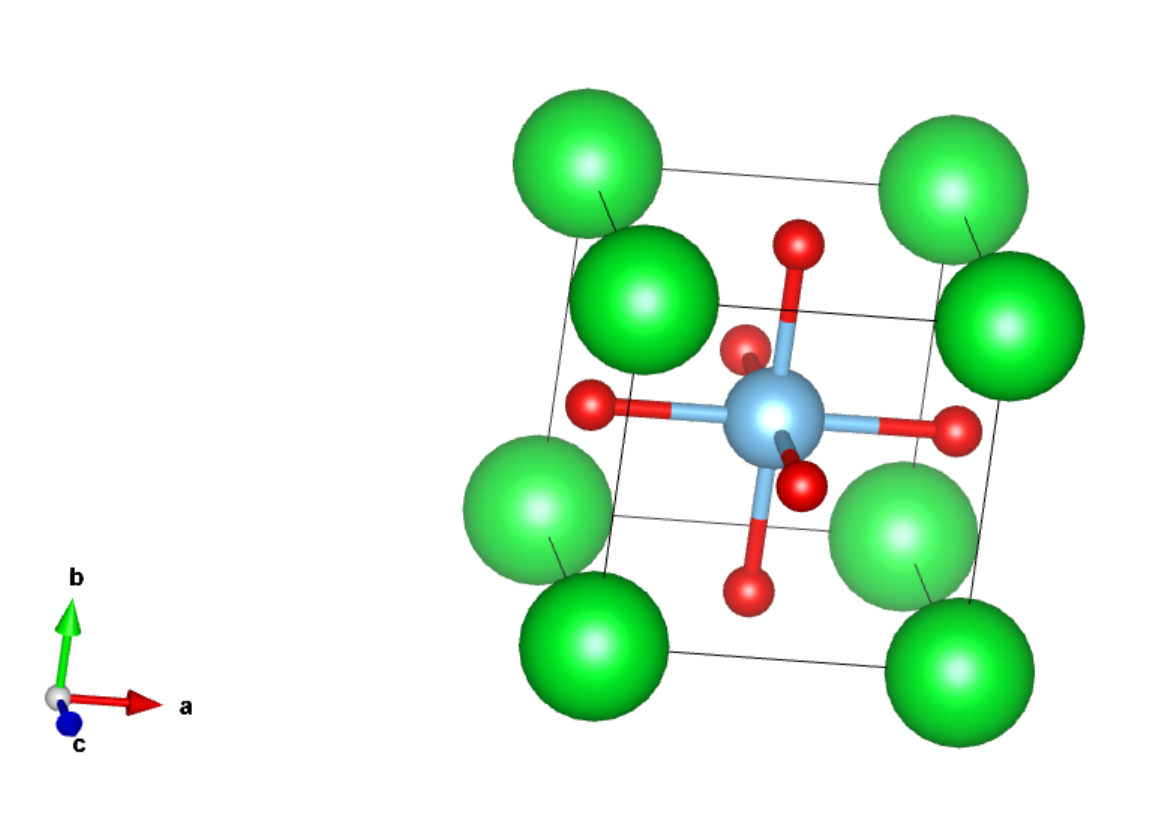}
        \caption{SrTiO\textsubscript{3}(Predicted)}
        \vspace{-3pt}
        \label{fig:SrTiO3_predict}
    \end{subfigure}\hfill
    \begin{subfigure}[t]{0.33\textwidth}
        \includegraphics[width=\textwidth]{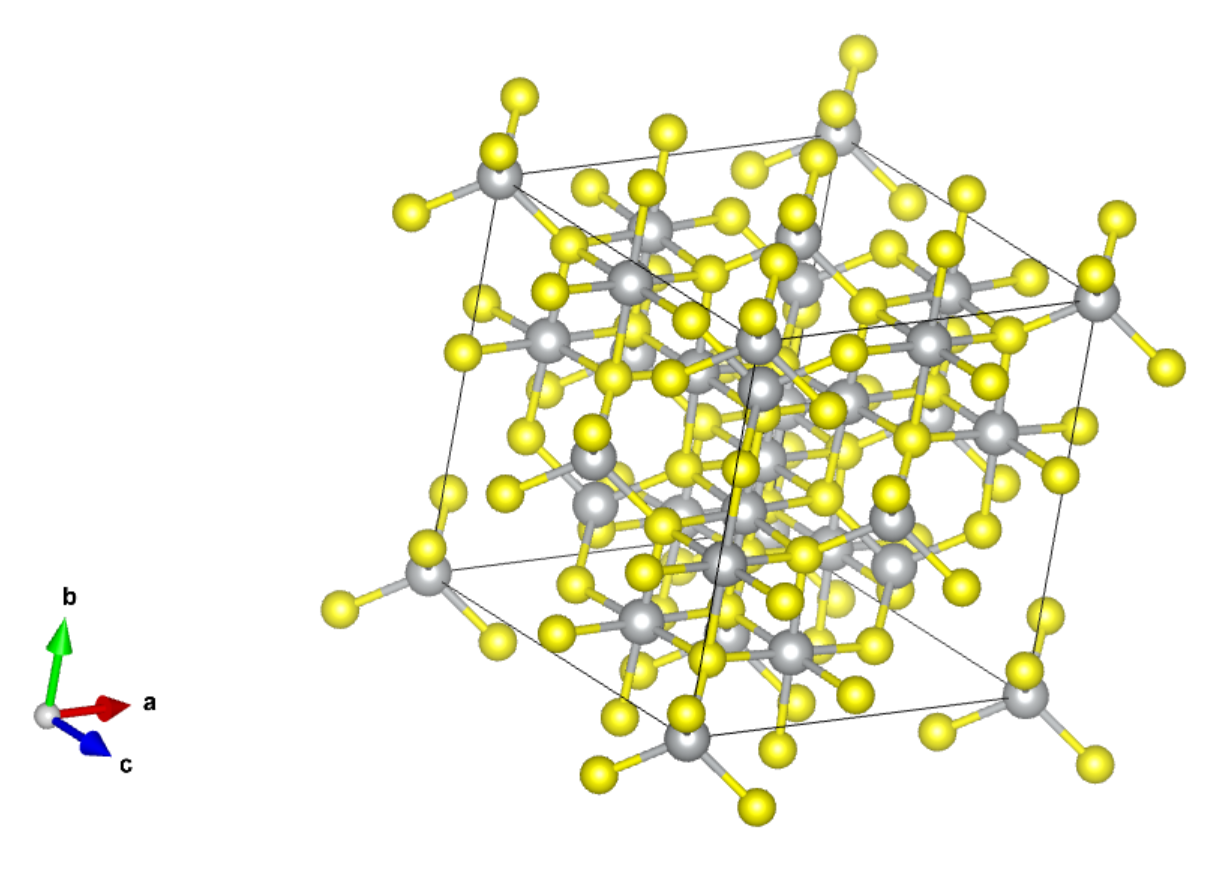}
        \caption{Ni\textsubscript{3}S\textsubscript{4}(Predicted)}
        \vspace{-3pt}
        \label{fig:Ni3S4_predict}
    \end{subfigure}
    \begin{subfigure}[t]{0.33\textwidth}
        \includegraphics[width=\textwidth]{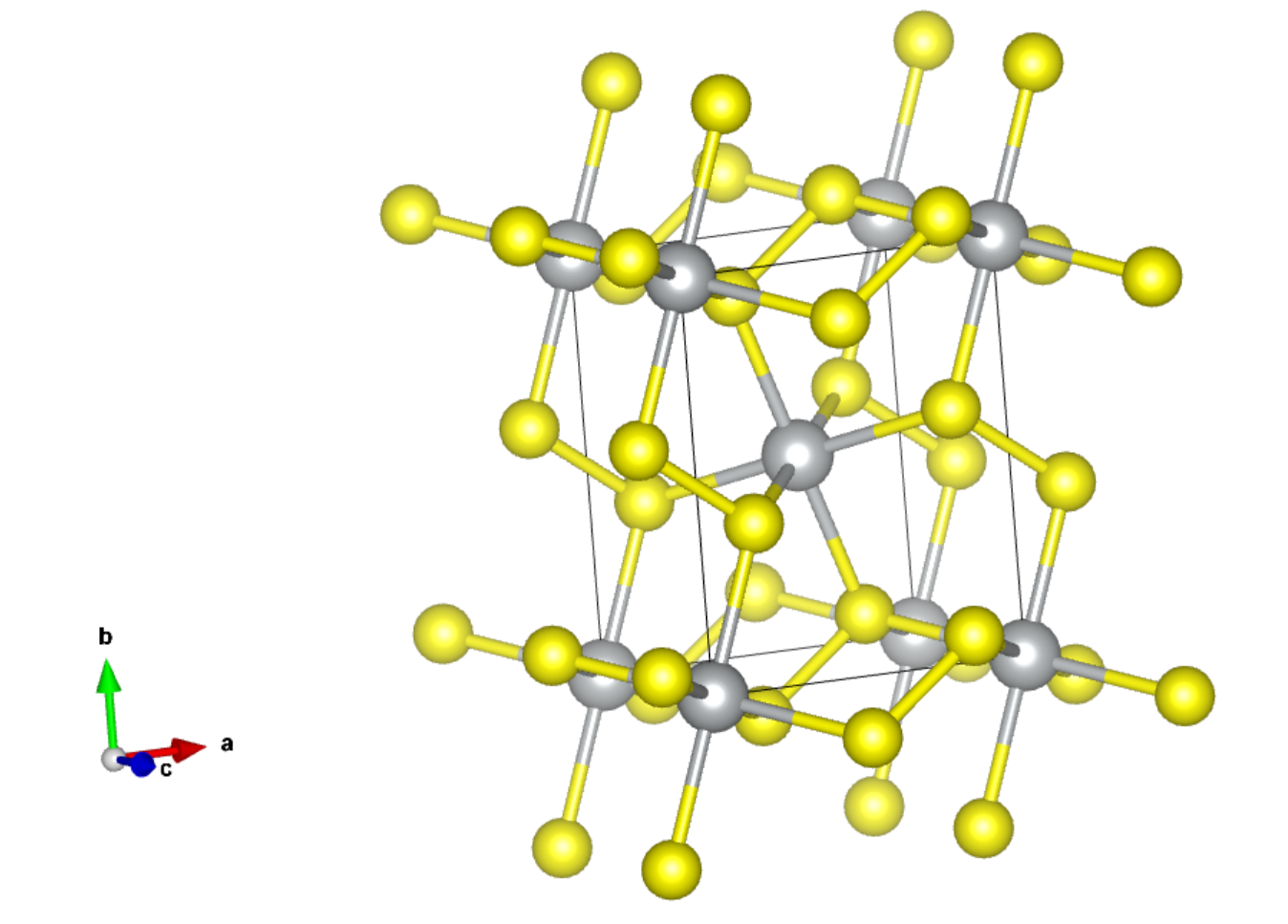}
        \caption{NiS\textsubscript{2}(Predicted)}
        \vspace{-3pt}
        \label{fig:NiS2_predict}
    \end{subfigure}\hfill
 \begin{subfigure}[t]{0.33\textwidth}
        \includegraphics[width=\textwidth]{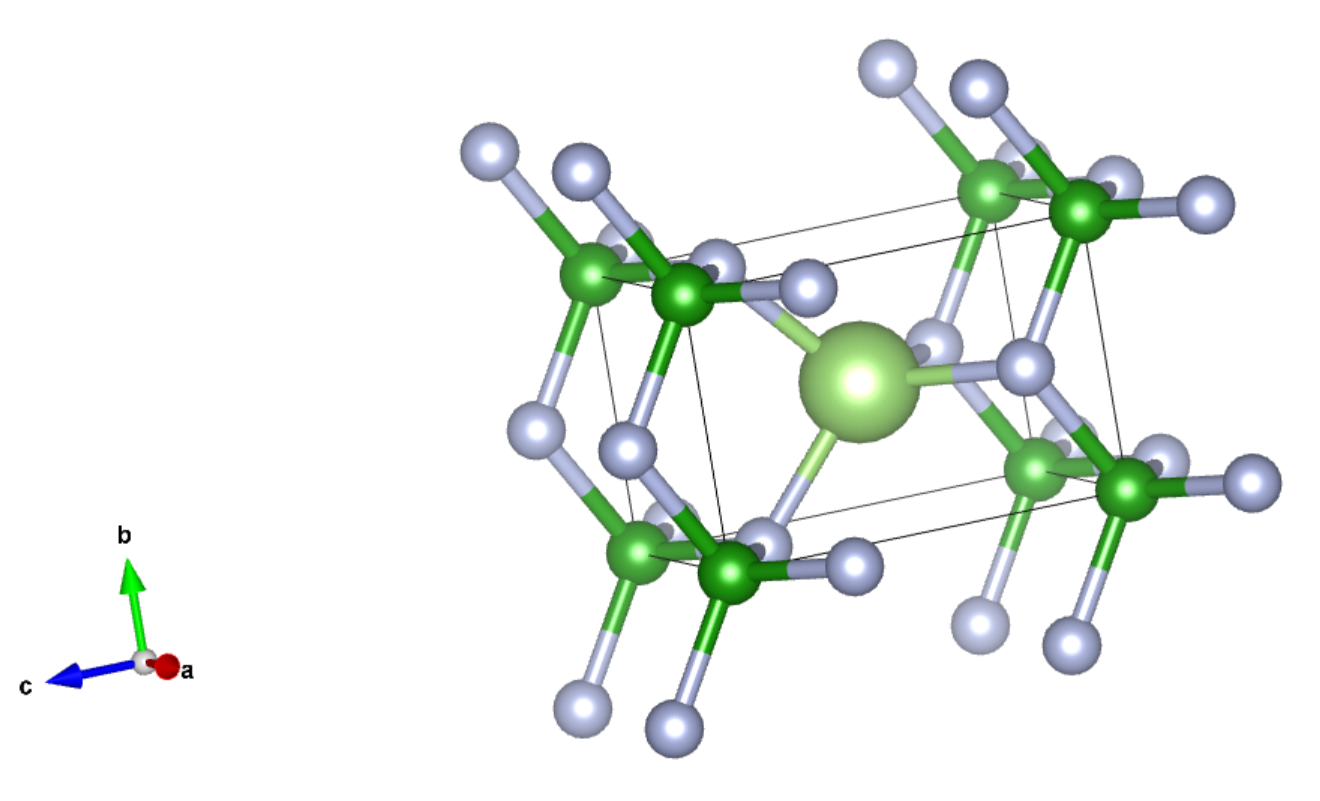}
        \caption{GaBN\textsubscript{2}(Target)}
        \vspace{-3pt}
        \label{fig:GaBN2_target}
    \end{subfigure}
    \begin{subfigure}[t]{0.33\textwidth}
        \includegraphics[width=\textwidth]{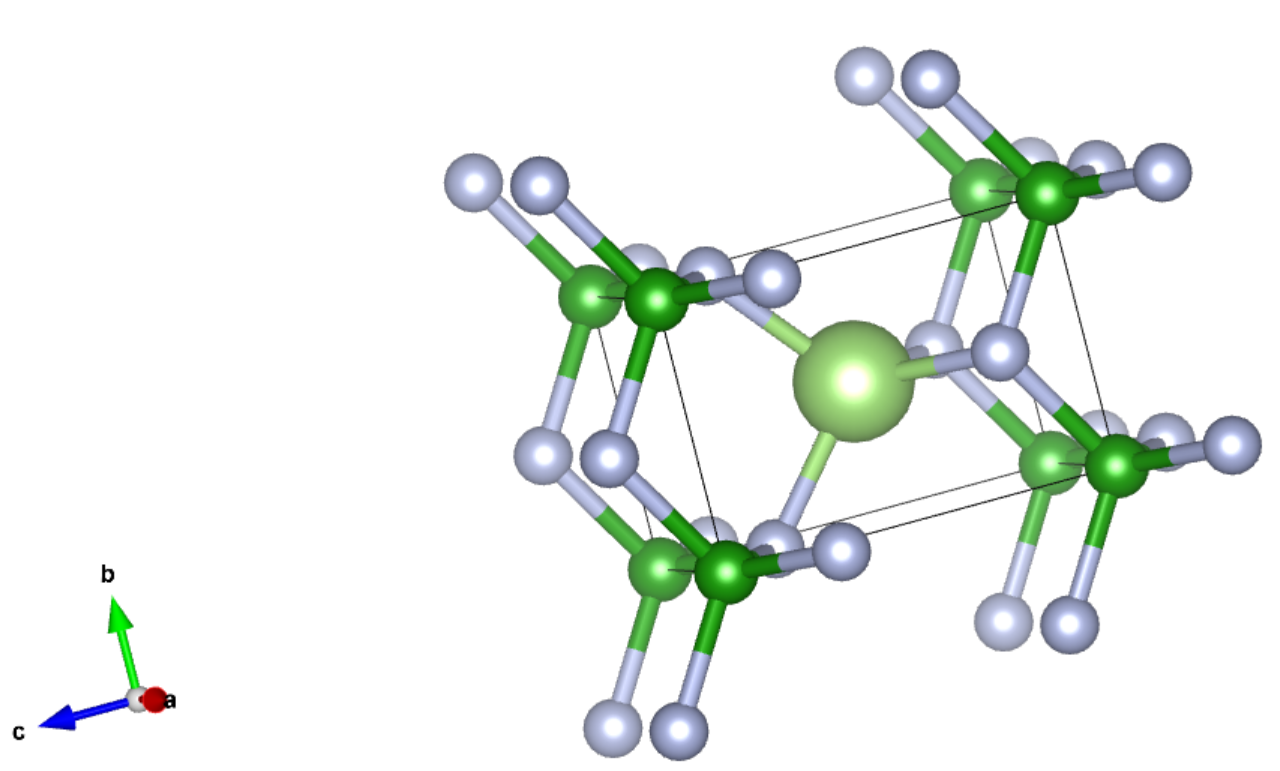}
        \caption{GaBN\textsubscript{2}(Predicted1)}
        \vspace{-3pt}
        \label{fig:GaBN2_predict1}
    \end{subfigure}\hfill    
    \begin{subfigure}[t]{0.33\textwidth}
        \includegraphics[width=0.9\textwidth]{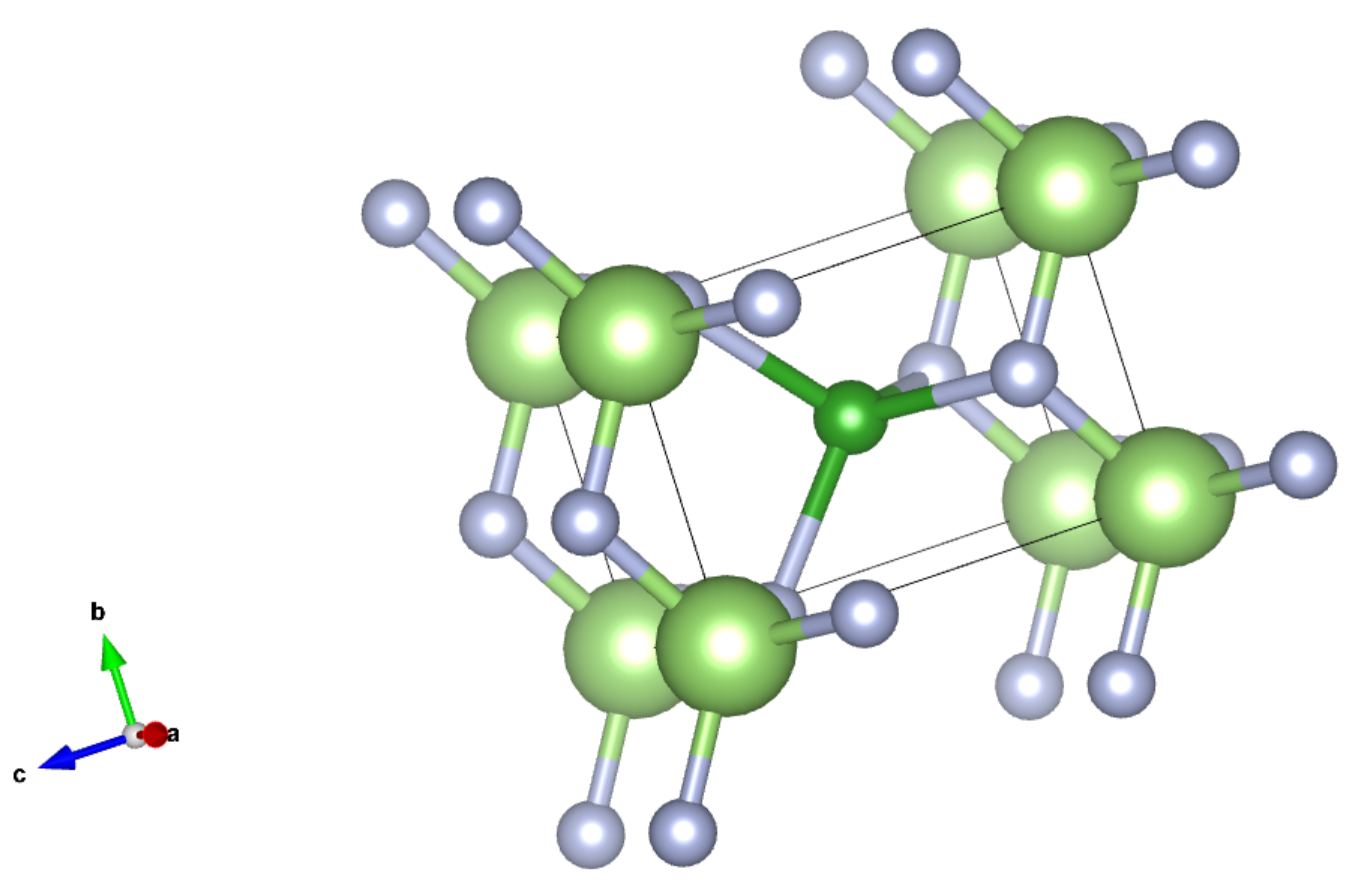}
        \caption{GaBN\textsubscript{2}(Predicted2)}
        \vspace{-3pt}
        \label{fig:GaBN2_predict2}
    \end{subfigure}\hfill    
    \begin{subfigure}[t]{0.33\textwidth}
        \includegraphics[width=\textwidth]{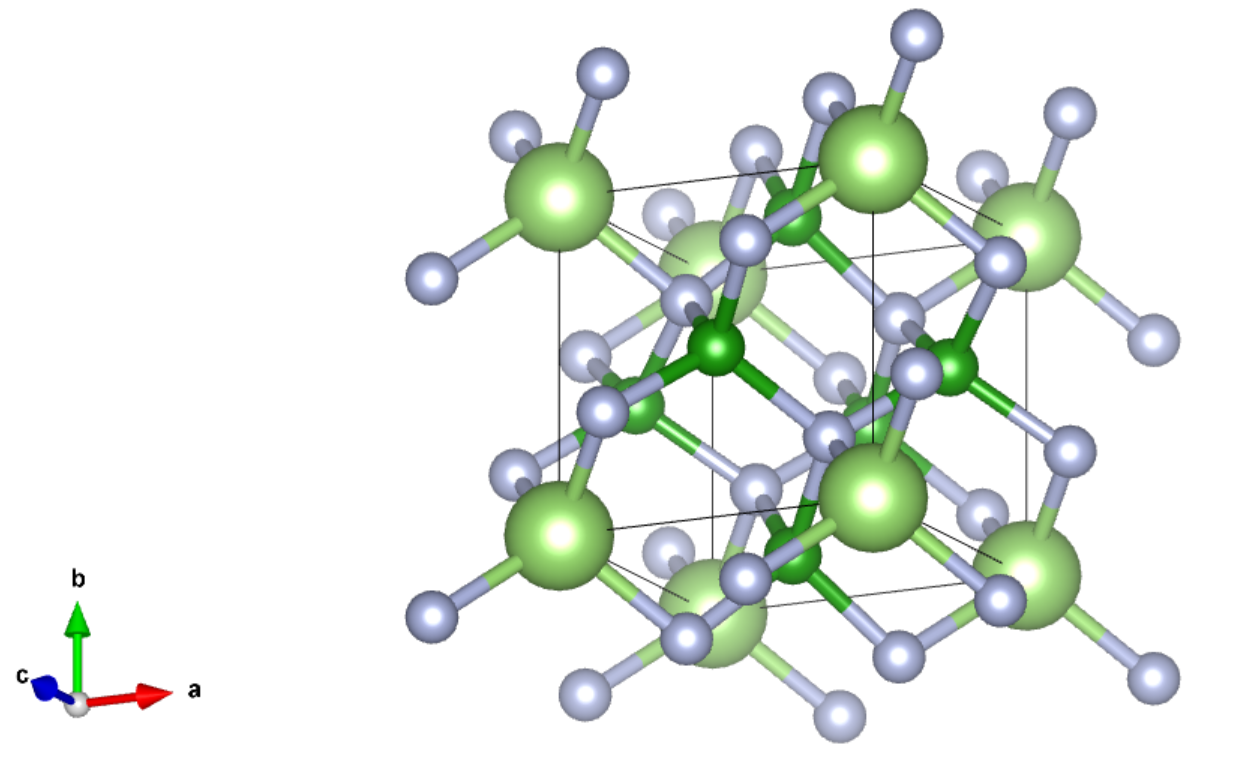}
        \caption{GaB\textsubscript{3}N\textsubscript{4}(Target)}
        \vspace{-3pt}
        \label{fig:GaB3N4_target}
    \end{subfigure}    
        \begin{subfigure}[t]{0.33\textwidth}
        \includegraphics[width=\textwidth]{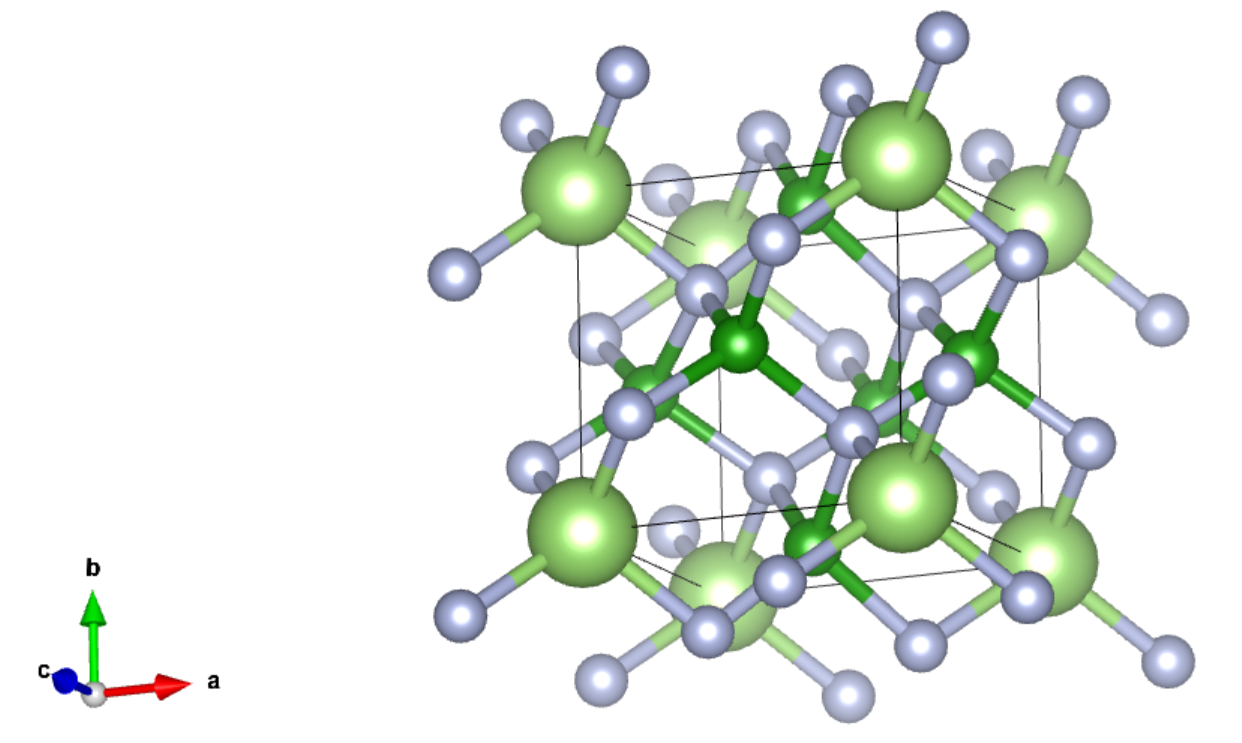}
        \caption{GaB\textsubscript{3}N\textsubscript{4}(Predicted1)}
        \vspace{-3pt}
        \label{fig:GaB3N4_predict1}
    \end{subfigure}\hfill
    \begin{subfigure}[t]{0.33\textwidth}
        \includegraphics[width=\textwidth]{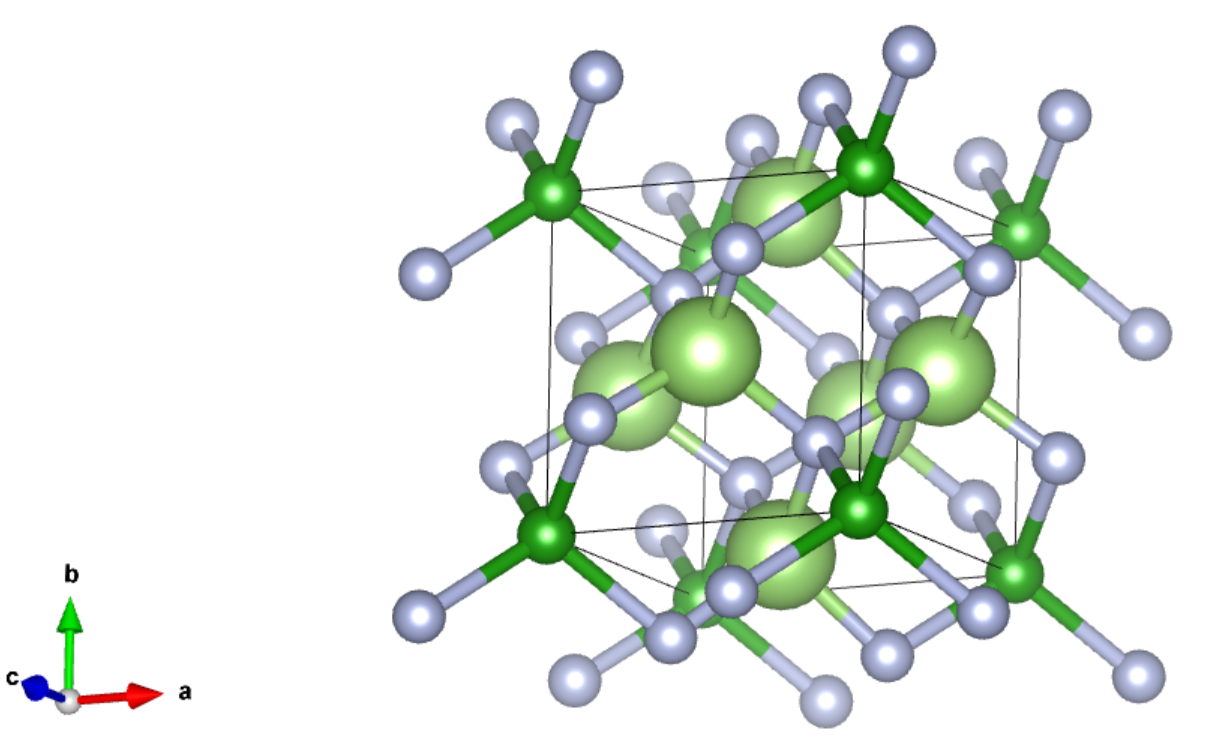}
        \caption{GaB\textsubscript{3}N\textsubscript{4}(Predicted2)}
        \vspace{-3pt}
        \label{fig:GaB3N4_predict2}
    \end{subfigure}
    \begin{subfigure}[t]{0.33\textwidth}
        \includegraphics[width=\textwidth]{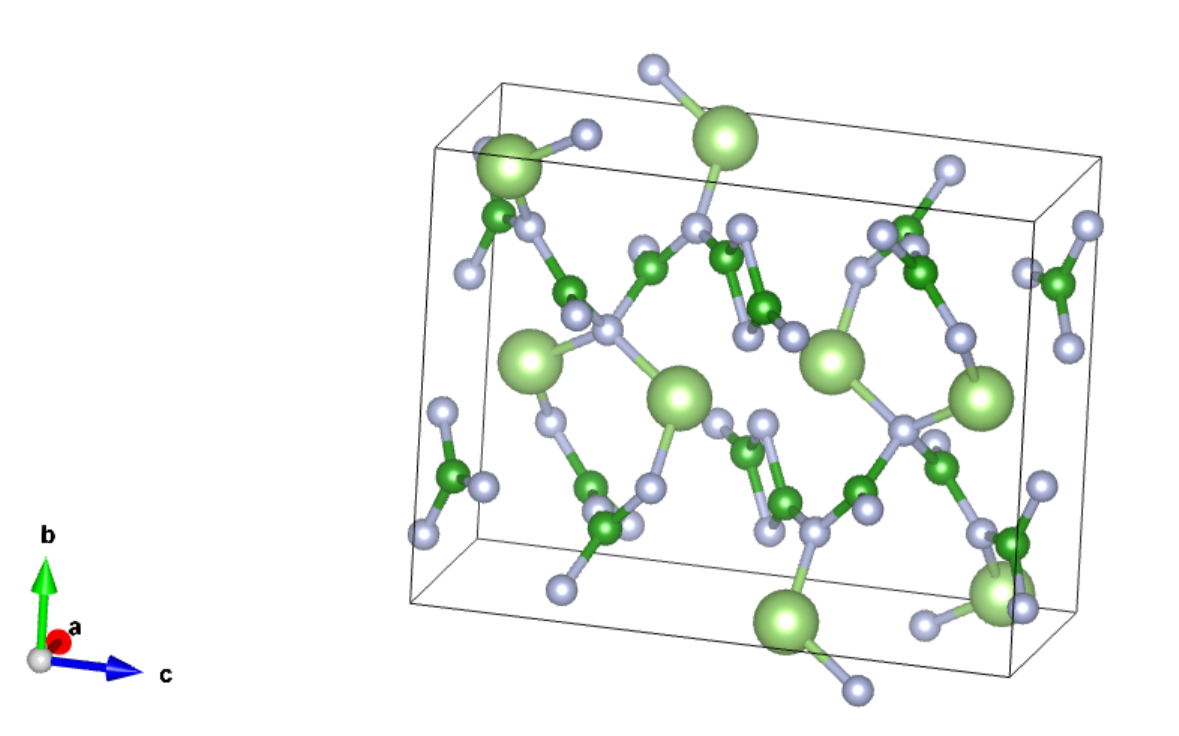}
        \caption{GaB\textsubscript{2}N\textsubscript{3}(Target)}
        \vspace{-3pt}
        \label{fig:GaB2N3_target}
    \end{subfigure}          
    \begin{subfigure}[t]{0.33\textwidth}
        \includegraphics[width=\textwidth]{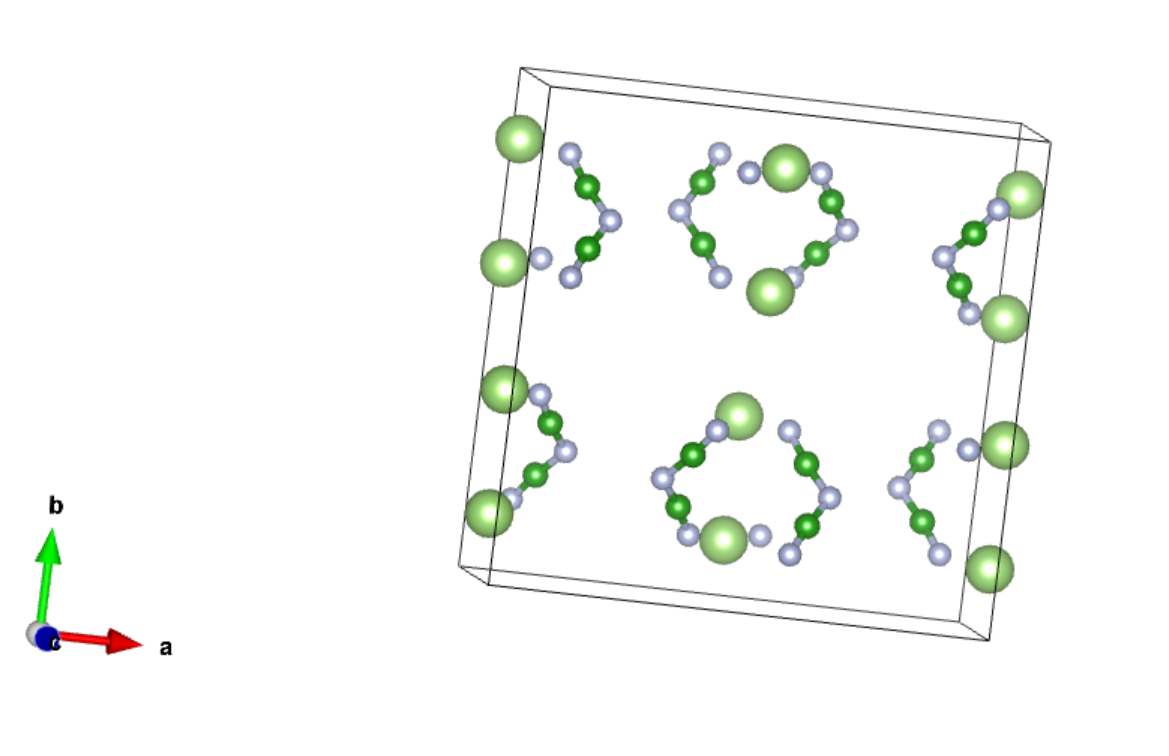}
        \caption{GaB\textsubscript{2}N\textsubscript{3}(Predicted1)}
        \vspace{-3pt}
        \label{fig:GaB2N3_predict1}
    \end{subfigure} 
    \begin{subfigure}[t]{0.33\textwidth}
        \includegraphics[width=\textwidth]{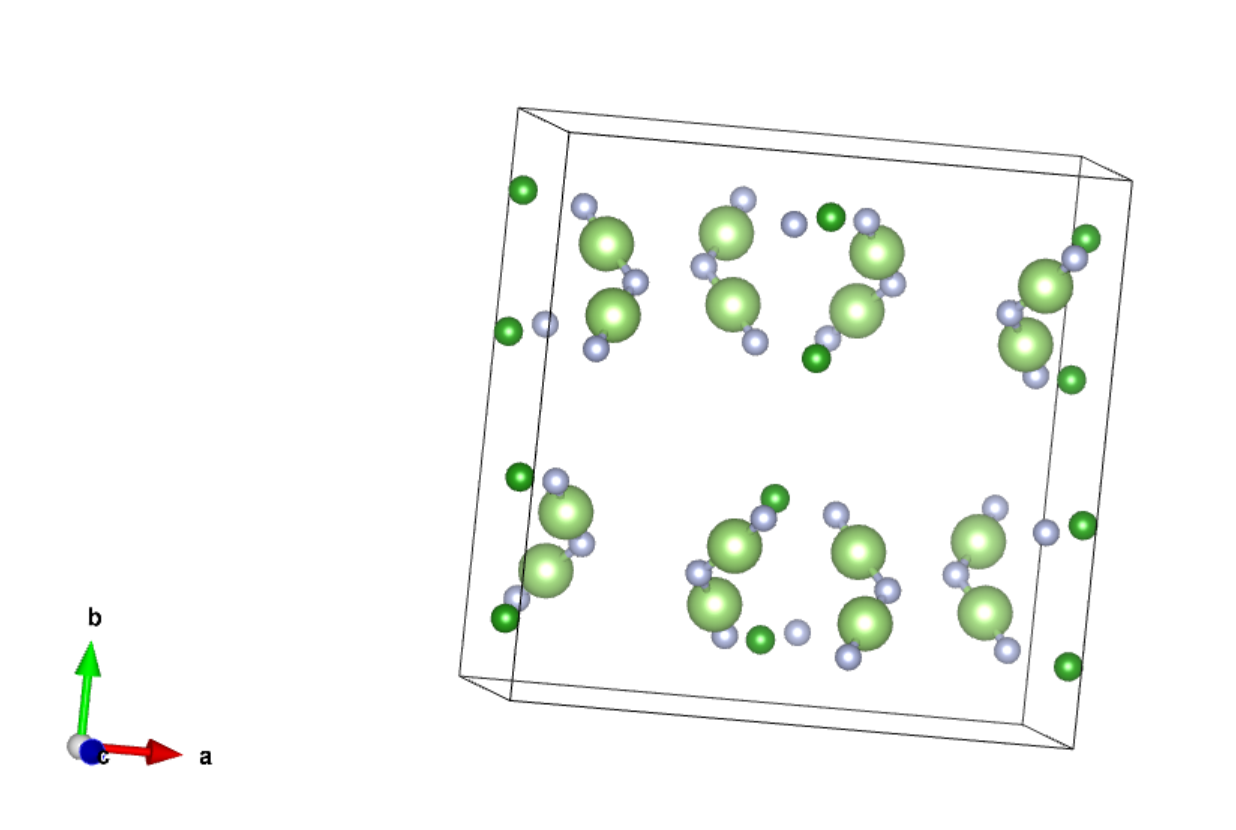}
        \caption{GaB\textsubscript{2}N\textsubscript{3}(Predicted2)}
        \vspace{-3pt}
        \label{fig:GaB2N3_predict2}
    \end{subfigure}     

   \caption{predicted structures by TCSP compared to targets.}
  \label{fig:predictedstructures}
\end{figure}

We selected seven formulas of target structures from materials project database for evaluating the capability of our TCSP algorithm for structure prediction. The first test target formula is SrTiO\textsubscript{3}, which has three different phases corresponding to space groups of 140, 149, 221. Its most famous structure is the cubic perovskite structure as shown in Figure \ref{fig:SrTiO3_target}. Our algorithm identified thousands of compatible templates and picked top 10 as templates including BaZrO\textsubscript{3} (mp-3834), MgTiO\textsubscript{3} (mp-1016830), CaZrO\textsubscript{3} (mp-542112), MgZrO\textsubscript{3} (mp-1017000), BaTiO\textsubscript{3} (mp-504715), BaTiO\textsubscript{3} (mp-5020), SrZrO\textsubscript{3} (mp-613402), CaTiO\textsubscript{3} (mp-5827), BaTiO\textsubscript{3} (mp-2998), SrHfO\textsubscript{3} (mp-4551), among which all are cubic templates except BaTiO\textsubscript{3}(mp-5020), which is a trigonal structure with space group 160. The top four predicted structures all have a zero rmsd error compared to the perovskite target structure: they all have the identical fractional coordinates as the target structure except that the cubic lengths are different (the predicted cubic structure in Figure \ref{fig:SrTiO3_predict} has a lattice length of 4.256 Å while the target structure has a lattice length of 3.945 Å), which may be fine-tuned using DFT based relaxation.

The second test sample is Ni\textsubscript{3}S\textsubscript{4}, which only has one cubic phase with the space group of 227. The structure is shown in Figure  \ref{fig:Ni3S4_target}. Our algorithm found top four templates including Co\textsubscript{3}S\textsubscript{4} (mp-943), Co\textsubscript{3}Se\textsubscript{4} (mp-20456), Ni\textsubscript{3}Se\textsubscript{4} (mp-1120781), Co\textsubscript{3}O\textsubscript{4} (mp-18748), all of them are cubic structures with the space group of 227. The predicted structure with the lowest rmsd is 0.000714  which is predicted by our algorithm using Co\textsubscript{3}S\textsubscript{4} as the template. The rmsd errors of the structures from the other three templates are much larger, all around 0.288 . We can see that the predicted structure in Figure \ref{fig:Ni3S4_predict} is very close to the target structure in Figure \ref{fig:Ni3S4_target}. We also found that the predicted structure of NiS\textsubscript{2} in Figure \ref{fig:NiS2_predict} also matches well with the target structure in Figure \ref{fig:NiS2_target}, which has the small rmsd error of 0.004918 .

We also tested four formulas of the chemical system Ga-B-N including GaBN\textsubscript{2}, GaB\textsubscript{3}N\textsubscript{4}, Ga\textsubscript{2}BN\textsubscript{3}, and Ga\textsubscript{3}BN\textsubscript{4}. For the GaBN\textsubscript{2} with the space group of 115, the best eight templates are found by our algorithm. Five of them including AlBN\textsubscript{2} (mp-1008557)  ,AlBN\textsubscript{2} (mp-1008557), AlGaP\textsubscript{2} (mp-1228888), AlGaN\textsubscript{2} (mp-1008556), B\textsubscript{2}AsP(mp-1008528) have the same tetragonal crystal system as the target structure with the space group of 115. The remaining are trigonal with the space group of 166. The predicted structures AlBN\textsubscript{2} (mp-1008557) and AlBN\textsubscript{2} (mp-1008557) have the same lowest rmsd error of 0.003889 . However they have different structure patterns as shown in Figure \ref{fig:GaBN2_predict1} and Figure \ref{fig:GaBN2_predict2}. For the same template, our algorithm suggests two element replacement strategies. In the first strategy, the element Ga in the test formula is used to replace the element Al in the template AlBN\textsubscript{2}; in the second strategy, the element Ga in the test formula GaBN\textsubscript{2} replaces the element B in the same template AlBN\textsubscript{2}. For the formula GaB\textsubscript{3}N\textsubscript{4} with the space group of 215, TCSP finds top 9 most similar templates AlB\textsubscript{3}N\textsubscript{4} (mp-1019379), AlB\textsubscript{3}N\textsubscript{4} (mp-1019379), CrGa\textsubscript{3}P\textsubscript{4} (mp-985440), AlGa\textsubscript{3}N\textsubscript{4} (mp-1019508), Al\textsubscript{3}GaN\textsubscript{4} (mp-1019378), Al\textsubscript{3}BN\textsubscript{4} (mp-1019380), Al\textsubscript{3}BN\textsubscript{4} (mp-1019380), Ga\textsubscript{3}BN\textsubscript{4} (mp-1019743), Ga\textsubscript{3}BN\textsubscript{4} (mp-1019743) of the same space group 215 as well. The lowest rmsd error with different structure templates AlB\textsubscript{3}N\textsubscript{4} and AlB\textsubscript{3}N\textsubscript{4} is 0.002336  by using two element replacement strategies. The element Ga in the first strategy is used to replace Al in the template as shown in Figure \ref{fig:GaB3N4_predict1} and
as shown in Figure \ref{fig:GaB3N4_predict2}, Ga replaces B in  AlB\textsubscript{3}N\textsubscript{4} in the second strategy. For the formula GaB\textsubscript{2}N\textsubscript{3} (mp-1245554) in Figure \ref{fig:GaB2N3_target} which has monoclinic structure with the space group 15, TCSP only find two templates AuC2N3 (mp-1245653) and AuC2N3 (mp-1245653) with the same space group. The lowest rmsd is 0.24746 for these two predicted structures. As shown in Figure \ref{fig:GaB2N3_predict1}, our algorithm uses the first strategy,  TCSP uses Ga in the test formula GaB\textsubscript{2}N\textsubscript{3} to replace Au in the first template AuC\textsubscript{2}N\textsubscript{3}. In the second strategy, Ga replaces B in the second template as shown in Figure \ref{fig:GaB2N3_predict2}. For the formula Ga\textsubscript{3}BN\textsubscript{4}(mp-1019743) with the space group 215, TCSP finds 9 most similar templates Al\textsubscript{3}BN\textsubscript{4} (mp-1019380), Al\textsubscript{3}BN\textsubscript{4} (mp-1019380), Al\textsubscript{3}GaN\textsubscript{4} (mp-1019378), AlGa\textsubscript{3}N\textsubscript{4} (mp-1019508), CrGa\textsubscript{3}P\textsubscript{4} (mp-985440), AlB\textsubscript{3}N\textsubscript{4} (mp-1019379), AlB\textsubscript{3}N\textsubscript{4} (mp-1019379), GaB\textsubscript{3}N\textsubscript{4} (mp-1019740), GaB\textsubscript{3}N\textsubscript{4} (mp-1019740) with the same space group 215. The templates Al\textsubscript{3}BN\textsubscript{4} (mp-1019380) and Al\textsubscript{3}BN\textsubscript{4} (mp-1019380) with different structure patterns have the same lowest rmsd error of 0.004017781. For Al\textsubscript{3}BN\textsubscript{4}, AlB\textsubscript{3}N\textsubscript{4} and GaB\textsubscript{3}N\textsubscript{4}, they all have different structure patterns by using the two elements replacement strategies.


Our TCSP algorithm can output multiple predictions using different templates. To understand this capability, Table~\ref{tab:results_materials} shows the RMSD and quality scores of the top 10 predictions for each input formula. For SrTiO\textsubscript{3}, the top 4 structures all have zero RMSD errors for their fractional coordinates with their replacement distance scores ranging from 2 to 3. The MAE errors of these four structures are also 0. We do find their lattice length is different from the target structures, which, however, can be tuned by DFT-based structure relaxation. For Ni\textsubscript{3}S\textsubscript{4}, only the top 1 result is very close to the target structure with three much worse results. For NiS\textsubscript{2}, the top 2 predicted structures have RMSD of 0.0049  and 0.0124. Both are good compared to the target structures. For GaBN\textsubscript{2}, the top four results all have small RMSD errors ranging from 0.0039 to 0.0209. The same is true for the predicted structures of GaB\textsubscript{3}N\textsubscript{4}. The worst prediction performance is on the formula GAB\textsubscript{2}N\textsubscript{3}, which can only find two compatible templates, both leading to very different structures from the target structure. Their MAE errors are 0.1853 . For GA\textsubscript{3}BN\textsubscript{4}, the top three predictions are all of high quality with RMSD of only 0.004/0.004/0.0206  respectively. In terms of the quality score distribution, we found that low replacement distance scores indicate predicted structures with good quality: for example, predictions with replacement distance scores greater than 5 are all low-quality results. However, low replacement distance scores do not always mean their structures are of high quality. For example, in the case of SrTiO3, the structures with high RMSD have lower replacement distance scores than those top 4 results.  

\FloatBarrier

\begin{table}[th]

\centering
\caption{Prediction performance (rmsd error) of top 10 results for each sample in the benchmark set by TCSP
}
\label{tab:results_materials}
\begin{tabular}{|llrrrrrrrrll|}
\hline
\rowcolor[HTML]{FFFFFF} 
\textbf{Formula}                                              & \textbf{Metric}                & \multicolumn{1}{l}{\cellcolor[HTML]{FFFFFF}\textbf{Top1}}    & \multicolumn{1}{l}{\cellcolor[HTML]{FFFFFF}\textbf{Top2}} & \multicolumn{1}{l}{\cellcolor[HTML]{FFFFFF}\textbf{Top3}} & \multicolumn{1}{l}{\cellcolor[HTML]{FFFFFF}\textbf{Top4}} & \multicolumn{1}{l}{\cellcolor[HTML]{FFFFFF}\textbf{Top5}} & \multicolumn{1}{l}{\cellcolor[HTML]{FFFFFF}\textbf{Top6}} & \multicolumn{1}{l}{\cellcolor[HTML]{FFFFFF}\textbf{Top7}} & \multicolumn{1}{l}{\cellcolor[HTML]{FFFFFF}\textbf{Top8}} & \textbf{Top9}                        & \textbf{Top10}                       \\ \hline
\multicolumn{1}{|l|}{\cellcolor[HTML]{FFFFFF}SrTiO\textsubscript{3}} & \multicolumn{1}{l|}{rmsd}  & \multicolumn{1}{r|}{\cellcolor[HTML]{DDDDDD}0}      & \multicolumn{1}{r|}{\cellcolor[HTML]{DDDDDD}0}                           & \multicolumn{1}{r|}{\cellcolor[HTML]{DDDDDD}0}                           & \multicolumn{1}{r|}{\cellcolor[HTML]{DDDDDD}0}                           & \multicolumn{1}{r|}{0.1667}                      & \multicolumn{1}{r|}{0.2832}                      & \multicolumn{1}{r|}{0.4082}                      & \multicolumn{1}{r|}{0.4410}                      & \multicolumn{1}{r|}{0.4410} & \multicolumn{1}{r|}{0.4410} \\ \hline
\multicolumn{1}{|l|}{}                               & \multicolumn{1}{l|}{score} & \multicolumn{1}{r|}{2}                              & \multicolumn{1}{r|}{2}                           & \multicolumn{1}{r|}{2}                           & \multicolumn{1}{r|}{3}                           & \multicolumn{1}{r|}{1}                           & \multicolumn{1}{r|}{1}                           & \multicolumn{1}{r|}{1}                           & \multicolumn{1}{r|}{1}                           & \multicolumn{1}{r|}{1}      & \multicolumn{1}{r|}{2}      \\ \hline
\multicolumn{1}{|l|}{\cellcolor[HTML]{FFFFFF}Ni\textsubscript{3}S\textsubscript{4}}  & \multicolumn{1}{l|}{rmsd}  & \multicolumn{1}{r|}{\cellcolor[HTML]{DDDDDD}0.0007} & \multicolumn{1}{r|}{0.2885}                      & \multicolumn{1}{r|}{0.2888}                      & \multicolumn{1}{r|}{0.2897}                      & \multicolumn{1}{l|}{}                            & \multicolumn{1}{l|}{}                            & \multicolumn{1}{l|}{}                            & \multicolumn{1}{l|}{}                            & \multicolumn{1}{l|}{}       & \multicolumn{1}{l|}{}       \\ \hline
\multicolumn{1}{|l|}{\cellcolor[HTML]{FFFFFF}}       & \multicolumn{1}{l|}{score} & \multicolumn{1}{r|}{3}                              & \multicolumn{1}{r|}{4}                           & \multicolumn{1}{r|}{1}                           & \multicolumn{1}{r|}{4}                           & \multicolumn{1}{l|}{}                            & \multicolumn{1}{l|}{}                            & \multicolumn{1}{l|}{}                            & \multicolumn{1}{l|}{}                            & \multicolumn{1}{l|}{}       & \multicolumn{1}{l|}{}       \\ \hline
\multicolumn{1}{|l|}{\cellcolor[HTML]{FFFFFF}NiS\textsubscript{2}}   & \multicolumn{1}{l|}{rmsd}  & \multicolumn{1}{r|}{\cellcolor[HTML]{DDDDDD}0.0049} & \multicolumn{1}{r|}{0.0124}                      & \multicolumn{1}{r|}{0.0777}                      & \multicolumn{1}{r|}{0.2282}                      & \multicolumn{1}{r|}{0.2424}                      & \multicolumn{1}{r|}{0.2846}                      & \multicolumn{1}{l|}{}                            & \multicolumn{1}{l|}{}                            & \multicolumn{1}{l|}{}       & \multicolumn{1}{l|}{}       \\ \hline
\multicolumn{1}{|l|}{\cellcolor[HTML]{FFFFFF}}       & \multicolumn{1}{l|}{score} & \multicolumn{1}{r|}{3}                              & \multicolumn{1}{r|}{1}                           & \multicolumn{1}{r|}{3}                           & \multicolumn{1}{r|}{2}                           & \multicolumn{1}{r|}{1}                           & \multicolumn{1}{r|}{2}                           & \multicolumn{1}{l|}{}                            & \multicolumn{1}{l|}{}                            & \multicolumn{1}{l|}{}       & \multicolumn{1}{l|}{}       \\ \hline
\multicolumn{1}{|l|}{\cellcolor[HTML]{FFFFFF}GaBN\textsubscript{2}}  & \multicolumn{1}{l|}{rmsd}  & \multicolumn{1}{r|}{\cellcolor[HTML]{DDDDDD}0.0039} & \multicolumn{1}{r|}{\cellcolor[HTML]{DDDDDD}0.0039}                      & \multicolumn{1}{r|}{0.0174}                      & \multicolumn{1}{r|}{0.0209}                      & \multicolumn{1}{r|}{0.3410}                      & \multicolumn{1}{r|}{0.3412}                      & \multicolumn{1}{r|}{0.3412}                      & \multicolumn{1}{r|}{0.3886}                      & \multicolumn{1}{l|}{}       & \multicolumn{1}{l|}{}       \\ \hline
\multicolumn{1}{|l|}{\cellcolor[HTML]{FFFFFF}}       & \multicolumn{1}{l|}{score} & \multicolumn{1}{r|}{3}                              & \multicolumn{1}{r|}{1}                           & \multicolumn{1}{r|}{2}                           & \multicolumn{1}{r|}{1}                           & \multicolumn{1}{r|}{9}                           & \multicolumn{1}{r|}{1}                           & \multicolumn{1}{r|}{5}                           & \multicolumn{1}{r|}{31}                          & \multicolumn{1}{l|}{}       & \multicolumn{1}{l|}{}       \\ \hline
\multicolumn{1}{|l|}{GaB\textsubscript{3}N\textsubscript{4}}                         & \multicolumn{1}{l|}{rmsd}  & \multicolumn{1}{r|}{\cellcolor[HTML]{DDDDDD}0.0023} & \multicolumn{1}{r|}{\cellcolor[HTML]{DDDDDD}0.0023}                      & \multicolumn{1}{r|}{0.0152}                      & \multicolumn{1}{r|}{0.0162}                      & \multicolumn{1}{r|}{0.3335}                      & \multicolumn{1}{r|}{0.3344}                      & \multicolumn{1}{r|}{0.3344}                      & \multicolumn{1}{r|}{0.3347}                      & \multicolumn{1}{r|}{0.3347} & \multicolumn{1}{l|}{}       \\ \hline
\multicolumn{1}{|l|}{}                               & \multicolumn{1}{l|}{score} & \multicolumn{1}{r|}{1}                              & \multicolumn{1}{r|}{3}                           & \multicolumn{1}{r|}{22}                          & \multicolumn{1}{r|}{1}                           & \multicolumn{1}{r|}{1}                           & \multicolumn{1}{r|}{1}                           & \multicolumn{1}{r|}{3}                           & \multicolumn{1}{r|}{4}                           & \multicolumn{1}{r|}{0}      & \multicolumn{1}{l|}{}       \\ \hline
\multicolumn{1}{|l|}{\cellcolor[HTML]{FFFFFF}GaB\textsubscript{2}N\textsubscript{3}} & \multicolumn{1}{l|}{rmsd}  & \multicolumn{1}{r|}{\cellcolor[HTML]{DDDDDD}0.2475} & \multicolumn{1}{r|}{\cellcolor[HTML]{DDDDDD}0.2475}                      & \multicolumn{1}{l|}{}                            & \multicolumn{1}{l|}{}                            & \multicolumn{1}{l|}{}                            & \multicolumn{1}{l|}{}                            & \multicolumn{1}{l|}{}                            & \multicolumn{1}{l|}{}                            & \multicolumn{1}{l|}{}       & \multicolumn{1}{l|}{}       \\ \hline
\multicolumn{1}{|l|}{\cellcolor[HTML]{FFFFFF}}       & \multicolumn{1}{l|}{score} & \multicolumn{1}{r|}{11}                             & \multicolumn{1}{r|}{7}                           & \multicolumn{1}{l|}{}                            & \multicolumn{1}{l|}{}                            & \multicolumn{1}{l|}{}                            & \multicolumn{1}{l|}{}                            & \multicolumn{1}{l|}{}                            & \multicolumn{1}{l|}{}                            & \multicolumn{1}{l|}{}       & \multicolumn{1}{l|}{}       \\ \hline
\multicolumn{1}{|l|}{\cellcolor[HTML]{FFFFFF}Ga\textsubscript{3}BN\textsubscript{4}} & \multicolumn{1}{l|}{rmsd}  & \multicolumn{1}{r|}{\cellcolor[HTML]{DDDDDD}0.0040} & \multicolumn{1}{r|}{\cellcolor[HTML]{DDDDDD}0.0040}                      & \multicolumn{1}{r|}{0.0206}                      & \multicolumn{1}{r|}{0.3336}                      & \multicolumn{1}{r|}{0.3337}                      & \multicolumn{1}{r|}{0.3345}                      & \multicolumn{1}{r|}{0.3345}                      & \multicolumn{1}{r|}{0.3347}                      & \multicolumn{1}{r|}{0.3347} & \multicolumn{1}{l|}{}       \\ \hline
\multicolumn{1}{|l|}{\cellcolor[HTML]{FFFFFF}}       & \multicolumn{1}{l|}{score} & \multicolumn{1}{r|}{1}                              & \multicolumn{1}{r|}{3}                           & \multicolumn{1}{r|}{1}                           & \multicolumn{1}{r|}{1}                           & \multicolumn{1}{r|}{22}                          & \multicolumn{1}{r|}{1}                           & \multicolumn{1}{r|}{3}                           & \multicolumn{1}{r|}{0}                           & \multicolumn{1}{r|}{4}      & \multicolumn{1}{l|}{}       \\ \hline
\end{tabular}
\end{table}

\begin{figure}[ht]
  \centering
  \includegraphics[width=0.55\linewidth]{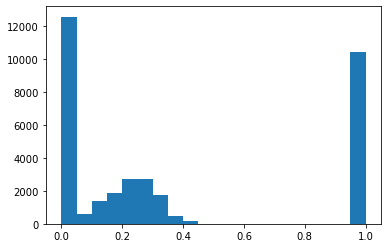}
  \caption{Distribution of rmsd errors of all Materials Project structures as predicted by our TCSP  using leave-one-out evaluation.}
  \label{fig:histogram_rmsd}
\end{figure}

To further evaluate the performance of our TCSP algorithm, we conduct comprehensive predictions of all 98,290 formulas in the Materials Project database using the leave-one-out evaluation approach. For each formula, we predict its structure using existing templates that do not have the same formula. Here we use the strict mode for finding templates: only top 10 templates with the same prototype and compatible oxidation states are used to predict new structures. For each formula, we first identify all its corresponding mp-ids with corresponding structures and then for each of these target structures, we pick the structure with the lowest rmsd error out of all the predicted structures and we show the distribution of these rmsd errors to see how our template based TCSP algorithm can recover the structures in the MP database. The result is shown in Figure \ref{fig:histogram_rmsd}. We find that for 34,569 MP structures, our algorithm has identified templates for structure prediction. Out of these target structures, TCSP has predicted hypothetical structures with a maximum RMSD less than 0.01 for 11,764 MP structures, or with RMDS less than 0.1 for 13,145 MP structures. We also find that for 10,433 structures, the algorithm could not find the correct templates that generates the same number of atoms in the unit cell for which we set the RMSD error at 1.0.

\FloatBarrier

\subsection{Discovery of new materials and DFT validation of the predict structures}

We are interested in how our TCSP algorithm can help to discover novel stable materials. We started with the Ga-B-N chemical system, for which we found four materials in the materials project database as shown in our test set discussed in the previous section. According to the materials project database, those available Ga-B-N structures have non-zero energy-above-hull values indicating those materials are thermodynamically unstable. We wonder if there exist thermodynamically and dynamically stable materials of this chemical system. We use the composition enumeration tool from our MaterialsAtlas.com toolbox to identify new Ga-B-N formulas prototypes and their predicted formation energy. We picked the top 41 formulas that do not exist in the MP database and used our TCSP to predict a set of candidate structures for each formula. We then conducted DFT relaxation and formation energy, e-above-hull energy, and phonon dispersion calculations to verify their thermodynamical and dynamical stability.

Our calculations (Table 2) show that almost all candidate structures found by our TCSP have negative formation energies. This is immensely helpful for discovering new materials using the first-principles calculations. If most of the candidate structures of a selected composition have positive formation energies, we have to waste computational hours to find the structures with negative formation energies. However, our ML model is able to filter out the unsuitable candidate structures to reduce the computational burden.

\begin{figure}[ht]
  \centering
  \includegraphics[width=0.8\linewidth]{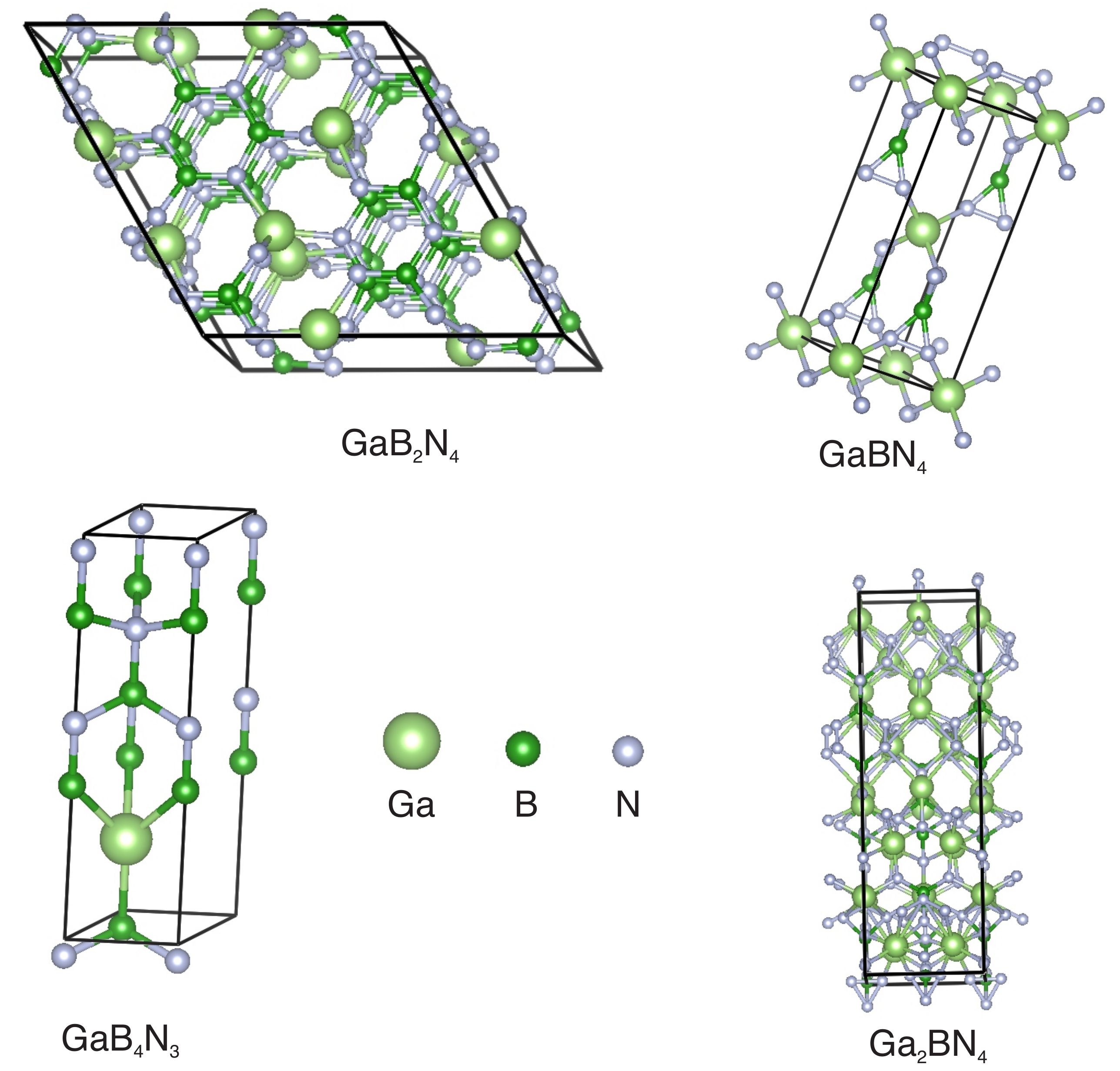}
  \caption{Candidate new structures with zero e-above-hull for GaB$_2$N$_4$ (using template mp-780282), GaB$_4$N$_3$ (using template mp-1224009), GaBN$_4$ (using template mp-30979), and Ga$_2$BN$_4$ (using template mp-698589).}
  \label{fig:struct}
\end{figure}

\begin{figure}[htb]
  \centering
   \begin{subfigure}[t]{0.49\textwidth}
        \includegraphics[width=\textwidth]{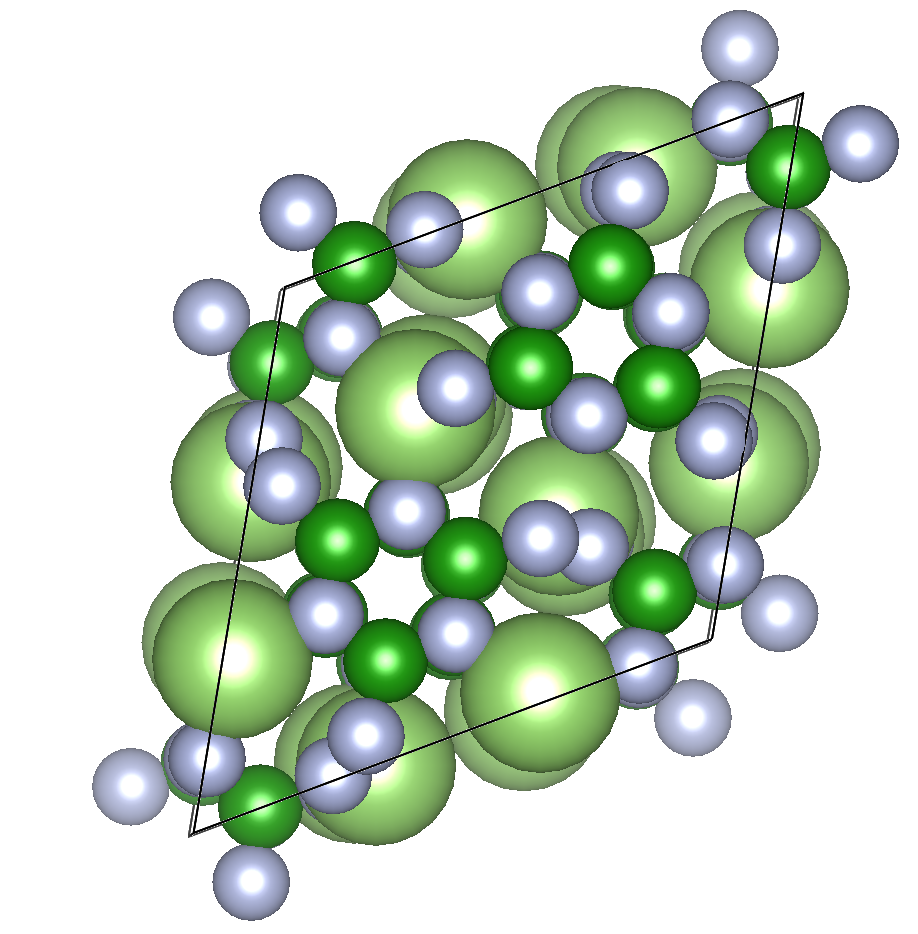}
        \caption{predicted structure of GaB$_2$N$_4$ }
        \vspace{-3pt}
        \label{fig:GaB2N3_predict1}
    \end{subfigure} 
     \begin{subfigure}[t]{0.5\textwidth}
        \includegraphics[width=\linewidth]{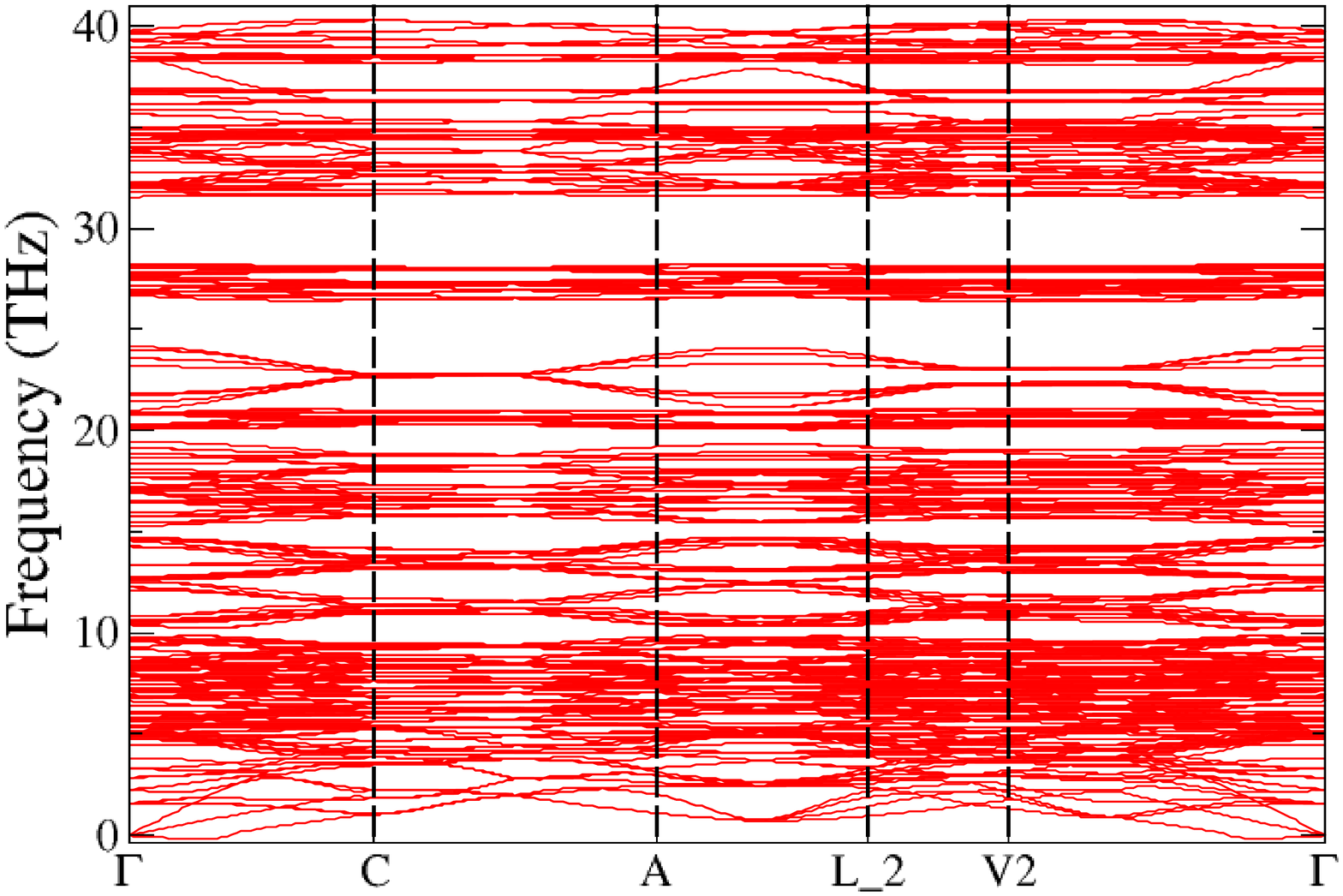}
          \caption{Phonon dispersion of GaB$_2$N$_4$ with mp-780282 template structure.}
          \label{fig:phonons}
    \end{subfigure} 
    \caption{New material GaB$_2$N$_4$ discovered by our TCSP with zero e-above-hull energy, negative formation energy (-3.2957 eV) and dynamical stability}
\end{figure}

We further calculated the e-above-hull to investigate the stability against the Ga-B-N competing phases. As given in the MP database, GaN, BN, Ga, B, and N2 are the stable competing phases available. Total energy calculations of competing phases were done with the same VASP settings used for Ga-B-N systems to determine the e-above-hull using the Pymatgen code. Our calculations suggest that 4 out of 41 materials exhibit zero e-bove-hull, indicating they are thermodynamically stable (Figure \ref{fig:struct}). Those materials and their candidate structures are shown in Table~2. We further carried out phonon calculations for the candidate structures with the lowest formation energies for those four materials. It is clear from Fig.~\ref{fig:phonons} that GaB$_2$N$_4$ material with mp-780282 template structure is dynamically stable at 0K temperature. It has an interesting layered honeycomb structure.

\begin{table}[]
\caption{Formation energy and corresponding templates of top 10 predictions for each of the four new materials}
\label{tab:my-table}
\begin{tabular}{|l|l|l|l|l|l|l|l|}
\hline
\multicolumn{2}{|c|}{GaB$_2$N$_4$}                                 & \multicolumn{2}{c|}{GaB$_4$N$_3$}                                   & \multicolumn{2}{c|}{Ga$_2$BN$_4$}                                  & \multicolumn{2}{c|}{GaBN$_4$}                                    \\ \hline
\multicolumn{1}{|c|}{mp-id} & \multicolumn{1}{c|}{E$_\textrm{form}$(eV)} & \multicolumn{1}{c|}{mp-id} & \multicolumn{1}{c|}{E$_\textrm{form}$ (eV)} & \multicolumn{1}{c|}{mp-id} & \multicolumn{1}{c|}{E$_\textrm{form}$(eV)} & \multicolumn{1}{c|}{mp-id} & \multicolumn{1}{c|}{E$_\textrm{form}$ (eV)} \\ \hline
mp-780282                   & -3.2957                        & mp-1224009                 & -2.4471                          & mp-532446                  & -2.9443                         & mp-30979                   & -3.29                            \\ \hline
mp-778103                   & -3.2414                        & mp-1225800                 & -2.2103                          & mp-698589                  & -2.8493                         & mp-20790                   & -2.9865                          \\ \hline
mp-13335                    & -3.2224                        & mp-1228436                 & -2.1092                          & mp-761314                  & -2.7696                         & mp-1224951                 & -2.7224                          \\ \hline
mp-780395                   & -3.1119                        & mp-1120750                 & -1.838                           & mp-5712                    & -2.5712                         & mp-1224810                 & -2.5462                          \\ \hline
mp-30161                    & -2.9124                        & mp-29672                   & -1.69                            & mp-1212041                 & -2.5703                         & mp-555538                  & -2.4165                          \\ \hline
mp-1194477                  & -2.4454                        & mp-1019378                 & -1.6865                          & mp-1255006                 & -2.5669                         & mp-1071955                 & -2.4012                          \\ \hline
mp-756317                   & -2.2645                        & mp-1228943                 & -0.3167                          & mp-765466                  & -2.5623                         & mp-1102285                 & -2.3984                          \\ \hline
mp-36866                    & -2.2639                        & mp-29672                   & -0.2741                          & mp-753397                  & -2.5622                         & mp-1071955                 & -2.398                           \\ \hline
mp-1208866                  & -2.1376                        & mp-1019508                 & -0.1409                          & mp-756649                  & -2.5621                         & mp-27462                   & -2.3971                          \\ \hline
N/A                            & N/A                                & mp-1223879                 & -0.0784                          & mp-1178203                 & -2.5621                         & mp-1102285                 & -2.3967                          \\ \hline
\end{tabular}
\end{table}


\section{Conclusion}

Crystal structure prediction plays a key role in new materials discovery \cite{oganov2019structure}. However, large-scale fast prediction of crystal structures is challenging, and user-friendly web apps are missing for such an important function despite the availability of a few public software that needs expensive high-performance computing (HPC) resources and expertise of computational materials. We believe such fast crystal structure prediction web apps are critical to the materials science community, as demonstrated by the bioinformatics field, which has more than 9000 web servers \cite{hu2021materialsatlas}. Here we propose a template-based crystal structure prediction algorithm, TCSP, and its companion web server for fast and quick crystal structure prediction. Due to the widely observed structure similarity across many materials families such as perovskites in the materials database, TCSP achieves strong prediction performance as benchmarked on the whole Materials Project structure using leave-one-out evaluation due to its flexible template selection algorithm using prototype and oxidation information. To our knowledge, this is the largest experiments for crystal structure prediction. We believe our web-based TCSP algorithm will be of great interest to materials scientists for exploratory materials discovery. 

\section{Data Availability}

Dataset is downloaded from the Materials Project database website. 

\section{Contribution}
Conceptualization, J.H.; methodology,J.H., L.W., W.Y., E.S., N.F., S.O., R.D.; software, J.H., L.W.,W.Y., N.F. ; resources, J.H.; writing--original draft preparation, J.H., L.W., E.S.,N.F.,R.X.; writing--review and editing,  J.H, L.W.; visualization, J.H., L.W., E.S.; supervision, J.H.;  funding acquisition, J.H.

\section*{Acknowledgement}
The research reported in this work was supported in part by National Science Foundation under the grant and 1940099 and 1905775. The views, perspectives, and content do not necessarily represent the official views of the NSF. We thank Andrew Hughes for his help in proofreading the manuscript. 

\bibliographystyle{unsrt}  
\bibliography{references}

\end{document}